\documentclass[aps,amsfonts,nofootinbib,tightenlines,showpacs,superscriptaddress]{revtex4}

\usepackage{bm}
\usepackage{bbm}
\usepackage{graphicx}
\usepackage[normalem]{ulem}
\usepackage[usenames,dvipsnames]{color}
\usepackage{mathrsfs}
\usepackage{amsmath}

\newcommand{\Vek}[1]{{\boldsymbol#1}}
\newcommand{\vek}[1]{\mathbf{#1}}
\newcommand{\KK}{\mathsf{K}}
\newcommand{\fract}[2]{{\textstyle\frac{#1}{#2}}}

\begin{document}

\title{On the Casimir Energy of Frequency Dependent Interactions}

\author{N. Graham}
\affiliation{Department of Physics, Middlebury College
Middlebury, VT 05753, USA}

\author{M. Quandt}
\affiliation{Institute for Theoretical Physics, University of T\"ubingen,
D--72076 T\"ubingen, Germany}

\author{H. Weigel}
\affiliation{Physics Department, Stellenbosch University,
Matieland 7602, South Africa}

\begin{abstract}
Vacuum polarization (or Casimir) energies can be straightforwardly computed 
from scattering data for static field configurations whose interactions with 
the fluctuating field are frequency independent. In effective theories, however, 
such interactions are typically frequency dependent. As a consequence, the
relationship between scattering data and the Green's function is
modified, which may or may not induce additional contributions
to the vacuum polarization energy. We discuss several examples that
naturally include frequency dependent interactions: (i) scalar
electrodynamics with a static background potential, (ii) an effective
theory that emerges from integrating out a heavy degree of
freedom, and (iii) quantum electrodynamics coupled to a
frequency dependent dielectric material. In the latter case, we argue
that introducing dissipation as required by the Kramers--Kronig
relations requires the consideration of the Casimir energy within a
statistical mechanics formalism, while in the absence of dissipation
we can work entirely within field theory, using an alternative
formulation of the energy density.
\end{abstract}

\maketitle

\section{Introduction}

Vacuum polarization (or Casimir) energies arise when quantum fluctuations 
interact with localized) background fields such as solitons. These interactions 
change the zero--point energies of the fluctuations. Summing these changes 
gives a non--trivial contribution to the energy at order $\hbar$, or in the 
language of Feynman diagrams, at one loop. As is common for loop diagrams, 
ultra--violet divergences occur in this calculation. It is essential that 
these divergences are unambiguously canceled by fixed counterterms whose 
coefficients do not depend on the specification of the background. In the 
case of \emph{static} backgrounds whose interactions with quantum fields do
not involve the conjugate momenta of these fields, spectral methods are well 
known to accomplish this task \cite{Graham:2009zz}. This formalism is well 
founded in quantum field theory \cite{Graham:2002xq}. For completeness we 
will provide a more heuristic derivation in this introduction, based on 
techniques from quantum mechanics, and review the essentials of the quantum 
field theory derivation in section II. The main goal of this paper, however, 
is to explore the spectral method in cases where the interaction is 
\emph{frequency dependent}, and to discuss the possible modifications necessary 
to handle such cases.

As mentioned earlier, it is necessary to sum the changes in the zero--point 
energies due to the interaction with the background. There are two types of energy 
eigenstates that lead to zero point energies, (i) bound states with discrete 
eigenvalues $\omega_n$ and (ii) continuum states with the dispersion relation 
$\omega=\frac{k^2}{2m}$ or $\omega^2=k^2+m^2$ for non--relativistic and relativistic 
theories, respectively\footnote{As usual in quantum field theory, we work in 
units where $\hbar = c = 1$.}.  Here $k$ is the asymptotic wave--number of the 
quantum fluctuation and $m$ is its mass. Since there are no bound states (b.s.) 
without the background, the bound states contribute just 
$\frac{1}{2}\sum_{n={\rm b.s.}}\,|\omega_n|$ to the vacuum polarization energy. 
The continuum piece is often expressed as an integral over the frequency weighted 
by the change of the density of states, $\Delta \rho(k)$, caused by the interaction. 
In order to find $\Delta \rho(k)$, boundary conditions far away from the interaction 
regime are imposed to discretize and  count the states. Eventually this boundary
is sent to infinity\footnote{An alternative formulation based directly on the 
change of the eigenfrequencies with the same boundary conditions can be found in 
Ref.~\cite{Ra82}.}. Far away from the interaction regime, however, all information 
about the background is encoded in the phase shift $\delta(k)$ that the interaction
imprints on the quantum fluctuation. A typical boundary condition is then of the 
form ${\rm sin}(kL+\delta(k))=0$, where $L$ refers to the position of the boundary.
Counting the states is then as simple are demanding that $kL+\delta(k)=n(k)\pi$. 
The density of states is the derivative $\partial n(k)/\partial k$ giving 
$\pi\,\Delta \rho(k)= \partial\delta(k)/\partial k$ as $L\to\infty$. This relation 
between the scattering phase shift and the density of states is also known as
the Krein--Friedel--Lloyd formula; see Ref.~\cite{Faulkner:1977} and
references therein. Putting these results together yields the {\it phase
shift formula} for the computation of vacuum polarization energies
\begin{eqnarray}
E_{\rm vac}&=&\int \frac{dk}{2\pi}\sqrt{k^2+m^2} \, 
\frac{\partial}{\partial k} \delta(k) 
+\frac{1}{2}\sum_{n={\rm b.s.}}\,|\omega_n|+E_{\rm c.t.}\cr\cr
&=&\int \frac{dk}{2\pi}\sqrt{k^2+m^2} \, 
\left[\frac{\partial}{\partial k} \delta(k)\right]_N
+\frac{1}{2}\sum_{n={\rm b.s.}}\,|\omega_n|+E_{\rm FD}^{(N)}+E_{\rm c.t.}\,,
\label{eq:master}
\end{eqnarray}
which also includes the counterterm (c.t.) contribution. In the second equation we have 
indicated the subtraction of the first $N$ terms of the Born series from the phase shift. 
The Born series orders scattering data according to increasing powers of the background 
interaction and may be constructed by iterating the wave--equation. The same expansion 
can be alternatively formulated in terms of Feynman diagrams, and $E_{\rm FD}$ in 
Eq.~(\ref{eq:master}) adds back in exactly what the Born terms subtracted. If $N$ is 
large enough the momentum integral will be finite, and the sum 
$E_{\rm FD}^{(N)}+E_{\rm c.t.}$ can be regularized and renormalized by standard techniques 
from perturbative field theory.  We stress that the Born and Feynman expansions have solely 
been introduced to identify and eventually cancel the ultra--violet divergences. This does 
not imply that these series are good approximations to the (one--loop) vacuum polarization 
energy; they might, in fact, not even converge. The spectral method sketched above has 
proven to be a very versatile tool to compute vacuum energies in sufficiently symmetric 
static backgrounds. It has been extended successfully to a wide range of situations, 
from cosmic strings \cite{Graham:2011fw} over soliton stabilization \cite{Farhi:2000ws}
to interfaces and extra dimensions \cite{Graham:2001dy}. The Krein--Friedel--Lloyd formula 
has been the basis of many studies of quantum corrections to classical soliton energies; 
see for example one dimensional models in Refs.\cite{Dashen:1974ci,Gervais:1975pa,
Cahill:1976im,Nakahara:1982,Graham:1998qq,FloresHidalgo:2002at} or the three dimensional Skyrme 
model in Refs.\cite{Moussallam:1991rj,Meier:1996ng,Weigel:2008zz}.

The above heuristic derivation shows that in order to find the density of states and 
thus the vacuum polarization energy, we must compare modes with different (albeit 
infinitesimally close) frequencies. Thus it is not at all obvious that the above 
argument for the density of states carries through when the interaction with the 
background becomes frequency dependent. This question arises irrespective of whether 
or not the field equations are diagonal in the frequency. Furthermore we have assumed 
that each mode contributes exactly $\fract{1}{2}\omega$. 
Effective action arguments suggest that this assumption essentially boils
down to the statement that even with the frequency dependent interaction
included the energy of the static background is still the negative  of its
Lagrange function.
This statement can undergoes modifications when the
interaction involves a time derivative of the quantum field. Here we
will consider a variety of models with frequency dependent
interactions and study  whether and under what conditions the above simple 
description for the vacuum energy can be maintained.

In the next section we review the quantum field theory derivation of 
Eq.~(\ref{eq:master}) for static interactions. Subsequently we investigate the 
modifications necessary for a number of examples with frequency dependent 
interactions. The simplest model that leads to a frequency dependent 
interaction is the gauge invariant coupling of a scalar field to a static 
electromagnetic background. In section III we study this example and find 
that computing its vacuum polarization energy does not require any 
approximation and, furthermore, that Eq.~(\ref{eq:master}) holds without 
modification.

Frequency dependent interactions also emerge in effective theories that 
result from integrating out some of the degrees of freedom. Such a scenario 
is often encountered in particle physics. Typically, a local energy density 
and a scattering problem can only be formulated in the effective theory once 
certain approximations are performed. But even with such approximations, 
frequency dependence must be retained to account for renormalizability. In 
the specific example of section IV we will see that Eq.~(\ref{eq:master}) is 
also valid without modifications. However, it is essential to consistently 
approximate both the wave--equations and the energy density when constructing 
the effective theory. 

In both examples we will observe that indeed significant changes to the energy 
density and the scattering data occur when frequency dependent interaction are 
involved. These changes tend to cancel, so that Eq.~(\ref{eq:master}) remains 
correct in such cases. However, the necessary cancellations of non--standard 
contributions involving the background potential are very delicate, and it is 
by no means obvious that the use of Eq.~(\ref{eq:master}) without modifications 
will succeed in \emph{all} cases. In section V we will study the important 
problem of electrodynamics in a dielectric. For the Casimir problem to remain 
under control in the ultraviolet, we must have a frequency--dependent dielectric 
function. In order to obey the Kramers-Kronig relations, such a dielectric must 
include dissipation, which in the Casimir problem represents a coupling of 
electromagnetic fluctuations to fluctuations in the microscopic matter degrees 
of freedom.  Refs.~\cite{Rahi:2009hm,Kenneth06,Kenneth:2007jk} formulate the 
Casimir energy in this case as an ensemble average in \emph{statistical mechanics}. 
In this approach the standard expression for the energy in a dielectric is 
identified from a calculation of the free energy. When translated intro scattering
theory, it induces extra terms in Eq.~(\ref{eq:master}).  These terms can play 
an important role in determining the sign of the Casimir self--stress 
\cite{Graham:2013yza}. Alternatively, one can consider a dielectric without 
dissipation, in which case the problem can be formulated within \emph{field theory}, 
at the cost of violation of the Kramers--Kronig relations. In this case, one 
obtains a different expression for the (now conserved) energy, one that has been 
obtained previously in classical electromagnetism as an approximation at
frequencies where dissipation is negligible \cite{Landau}, and that has also 
been discussed in the context of Casimir calculations
\cite{Brevik:2008ry,Milton:2010yw,PhysRevA.81.033812,PhysRevA.84.053813}.
Notably, this form of the energy has the effect of exactly canceling the extra 
terms appearing in Eq.~(\ref{eq:master}). We summarize our conclusions in 
section VI, and in an appendix we comment on the construction of a conserved 
energy from a wave--equation with an arbitrary frequency dependence.

\section{Review of the Spectral Method}

We start with a brief review of how the phase shift
formula~(\ref{eq:master}) for the vacuum polarization energy can be
derived in the case that the interaction is \emph{frequency independent}. 
For simplicity, we only consider the anti-symmetric channel in $d=1$ space 
dimensions, which can also be used to describe $S$-wave scattering in $d=3$.
For further details of this computation and the generalization to higher 
dimensions and angular momenta, the reader is referred to appendix A of 
Ref.~\cite{Graham:2002xq}. For a relativistic\footnote{Differences for 
non--relativistic scattering theory merely emerge via the dispersion relation.} 
boson of mass $m$, the scattering of a fluctuation mode with frequency 
$\omega$ and momentum $k$ (related by $\omega = \sqrt{k^2 + m^2}$)
off a static background $V(x)$ is described by the wave--equation
\begin{equation}
\left[-\partial_x^2+V(x)\right]\psi_k(x)=k^2\psi(x)
\label{eq:int1}
\end{equation}
from which the Green's function can be obtained as
\begin{equation}
G(x,x^\prime,k)=\frac{\phi_k(x_{<})f_k(x_{>})}{F(k)}\,.
\label{eq:int2}
\end{equation}
Here $\phi_k(x)$ and $f_k(x)$ are the regular and Jost solutions to the 
wave--equation~(\ref{eq:int1}) that are defined by suitable boundary conditions at 
$x=0$ and $x\to\infty$, respectively \cite{Newton:1982qc,Chadan:1977pq}. The 
denominator contains the Jost function $F(k)\sim\lim_{x\to0}f_k(x)$
whose negative phase is the scattering phase shift $\delta(k)$. 
By construction, $F(k)$ has simple zeros at $k=i\kappa_n$, where 
$\kappa_n$ are the bound state wave--numbers.
From the wave--equation~(\ref{eq:int1}) we also find the Wronskian
\begin{equation}
W\left[f_k(R),\phi_q(R)\right]=F(k)+(k^2-q^2)\int_0^R dx\, f_k(x)\phi_q(x)\,,
\label{eq:int3}
\end{equation}
for wave--functions with different momenta. Obviously, this relation must be 
modified once the frequency dependence on the right hand side of 
Eq.~(\ref{eq:int1}) is altered. For the moment, we adhere to Eq.~(\ref{eq:int3}) 
and differentiate with respect to $k$, take the limit $q\to k$, and divide
by $F(k)$ to obtain an equation for the Green's function at coincident points
\begin{equation}
W\left[\frac{\partial f_k(R)}{\partial k},\phi_q(R)\right]=
\frac{\partial}{\partial k}\,{\rm ln}\left[F(k)\right]
+2k\int_0^R dx\, G(x,x,k)\,.
\label{eq:int4}
\end{equation}
After subtracting the non--interacting ($V\equiv0$) analog of this formula and taking 
the limit $R\to\infty$, the left hand side vanishes \cite{Graham:2002xq}.
As a consequence,
\begin{equation}
0=\frac{\partial}{\partial k}\,{\rm ln}\left|F(k)\right|
-i\frac{\partial}{\partial k} \delta(k)
+2k\int_0^\infty dx\, \left[G(x,x,k)-G^{(0)}(x,x,k)\right]\,.
\label{eq:int5}
\end{equation}
The vacuum energy density in such simple systems may be expressed as
\begin{equation}
u(x)=\int \frac{dk}{\pi}\sqrt{k^2+m^2} \,
k \,\mathsf{Im}\left[G(x,x,k)-G^{(0)}(x,x,k)\right]
+\sum_{n={\rm b.s.}}\frac{|\omega_n|}{2}|\psi_n(x)|^2+\ldots\,,
\label{eq:int6}
\end{equation}
since the poles in the imaginary part of the Green's function determine the
density of scattering states. The discrete sum runs over the bound state solutions 
to the wave--equation~(\ref{eq:int1}) and the ellipsis denotes total derivative 
terms and counterterm contributions. Integrating the spatial
coordinate and taking the imaginary part of Eq.~(\ref{eq:int5}) 
yields the vacuum energy
\begin{equation}
E_{\rm vac}=\int_0^\infty dx\, u(x)=
\int \frac{dk}{2\pi}\sqrt{k^2+m^2} \, 
\frac{\partial}{\partial k} \delta(k) 
+\frac{1}{2}\sum_{n={\rm b.s.}}\,|\omega_n|\,
\label{eq:int7}
\end{equation}
up to renormalization. This is the phase shift formula relating the vacuum 
polarization energy of the background field configuration to the phase shift 
for quantum mechanical scattering off that background. The step between 
Eqs.~(\ref{eq:int3}) and~(\ref{eq:int4}) in this derivation shows that 
we should expect modifications to that formula must be expected when there is 
a more complicated frequency dependence in the wave--equation.

In the above we have followed the standard line of argument that is based
on relating the (local) density of states to the imaginary part of the 
Green's function at coincident points. Next we show that Eq.~(\ref{eq:int7}) can 
also be obtained from arguments based entirely on scattering data. This approach
is advantageous because the construction of a Green's function may be difficult 
for frequency dependent problems, since we may not have an underlying completeness 
relation for the scattering solutions. Instead, we only assume that the 
frequency dependence of the interaction does not create any (further) singularities
in the upper half plane of complex momenta. The energy density will involve
$\mathsf{Im}\,\Gamma(x,x,k)$ with
\begin{equation}
\Gamma(x,y,k)=\int_0^\infty \frac{dq}{2\pi} 
\frac{\psi^\ast_q(x)\psi_q(y)}{(k+i\epsilon)^2-q^2}
+\mbox{bound state contribution}
\label{eq:gamma1}
\end{equation}
where 
\begin{equation}
\psi_q(x)=\frac{q}{F(q)}\,\phi_q(x) 
=\frac{2i}{F(q)}\left[F(q)f_{-q}(x)-F(-q)f_{q}(x)\right]
\label{eq:gamma2}
\end{equation}
is the physical scattering solution. We need to show that $\Gamma(x,y,k)$ equals the 
right hand side of Eq.~(\ref{eq:int2}). For small arguments the real regular solution 
is well defined, as is the Jost solution for large arguments. Thus a suitable 
parameterization of the product of the scattering solutions is ($F^\ast(q)=F(-q)$ for 
real $q$)
\begin{equation}
\psi^\ast_q(x)\psi_q(y)=\frac{2iq}{F(q)F(-q)}\phi_q(x)
\left[F(q)f_{-q}(y)-F(-q)f_q(y)\right]
\label{eq:gamma3}
\end{equation}
in the case that $x=r_{<}$ and $y=r_{>}$. Since $\phi_q(x)=\phi_{-q}(x)$ we can extend 
the $q$--integral in Eq.~(\ref{eq:gamma1}) to cover the entire real axis by changing 
$q\to-q$ in the first term on the right hand side of Eq.~(\ref{eq:gamma3})
\begin{equation}
\Gamma(x,y,k)=2i\int_{-\infty}^\infty \frac{dq}{2\pi}\, \frac{q}{F(q)}
\frac{\phi_q(r_{<})f_q(r_{>})}{(k+i\epsilon)^2-q^2}
+\mbox{bound state contribution} \,.
\label{eq:gamma4}
\end{equation}
With the assumption that no additional singularities emerge from the frequency dependent 
interaction, the poles at $q=k+i\epsilon$ will directly yield the right hand side of 
Eq.~(\ref{eq:int2}). The poles due to the roots of the Jost function will 
compensate for the bound state contribution, and the roots at $q=-(k+i\epsilon)$ do 
not contribute because we close the contour integral in the upper half of the 
complex $q$--plane. Thus $\Gamma(x,x',k)$ equals the right hand side of 
Eq.~(\ref{eq:int2}) and the derivation of the phase shift formula can be completed 
without requiring that $\Gamma$ is a Green's function.

\section{Scalar Electrodynamics}

Scalar electrodynamics with a static electric potential produces a frequency 
dependent background  for the scalar quantum fluctuations. The simplest case is 
in $d=1$ spatial dimension so that the gauge field is $A^\mu=(V(x),0)$, which 
also satisfies the Lorentz condition $\partial_\mu A^\mu=0$. The case $d=1$ 
also has the advantage that all divergences cancel in a gauge invariant regularization, 
but otherwise the number of space dimensions is irrelevant for the present study. 
The Lagrangian for a scalar field $\Phi$ with mass $M$ coupled to an 
electromagnetic background field is
\begin{equation}
\mathcal{L}=\left(D_\mu \Phi\right)^\ast
\left(D^\mu \Phi\right)-M^2\Phi^\ast\Phi
\qquad {\rm with} \qquad D_\mu=\partial_\mu+iA_\mu\,,
\label{eq:lag}
\end{equation}
which implies the effective action of the photon field $A_\mu$,
\begin{equation}
S[A_\mu]=i\,{\rm Tr}\,{\rm log}\left[
\partial^2+M^2-i\overleftarrow{\partial}_\mu A^\mu
+iA_\mu\partial^\mu -A_\mu A^\mu \right]
=i\,{\rm Tr}\,{\rm log}\left[
\partial^2+M^2-i\overleftarrow{\partial}_t V
+iV\partial_t -V^2 \right]\,.
\label{eq:action}
\end{equation}
Here and in the following, the arrow denotes the direction of differentiation. 
The symbol ${\rm Tr}$ describes a functional trace that also includes space--time 
integration. Of course, this functional trace is superficially divergent at large 
momenta, and we implicitly assume some gauge invariant regularization, such as
dimensional regularization.

As a first step, we want to use functional methods to establish a relationship 
between the action in Eq.~(\ref{eq:action}) and the quantum (or vacuum polarization) 
energy. To do so we start from the energy density operator
\begin{equation}
T^{00}(x)=\frac{\partial\mathcal{L}}{\partial\dot{\Phi}}\dot{\Phi}
+\frac{\partial\mathcal{L}}{\partial\dot{\Phi}^\ast}\dot{\Phi}^\ast
-\mathcal{L}
=\Phi^\ast
\left(\overleftarrow{\partial}_t\overrightarrow{\partial}_t
+\overleftarrow{\partial}_x\cdot\overrightarrow{\partial}_x
+M^2-V^2\right)\Phi\,.
\label{eq:t00}
\end{equation}
The corresponding quantum energy density becomes
\begin{equation}
\langle T^{00}(x)\rangle =-i\,{\rm Tr}\,\Big[
\left(\partial^2+m^2-i\overleftarrow{\partial}_t V
+iV\partial_t -A_\mu A^\mu \right)^{-1}
\delta^{(2)}(x-x_{\rm op}) 
\left(\overleftarrow{\partial}_t\overrightarrow{\partial}_t
+\overleftarrow{\partial}_x\cdot\overrightarrow{\partial}_x
+M^2-V^2\right)\Big]\,.
\label{eq:vac1}
\end{equation}
Notice that $x_{\rm op}$ denotes the position operator, while $x$ is the argument of 
the density on the left hand side. This expression simplifies because
the potential is static, $V=V(x)$, so that the temporal part of 
the functional trace can be written as a single frequency integral. 
Furthermore the vacuum polarization energy is obtained by integration over 
all space, which removes the $\delta$--function completely,
\begin{equation}
E_{\rm vac} =
-i\int \frac{d\omega}{2\pi}\,{\rm Tr}^\prime\,
\Big[\left(-\omega^2-\partial_x^2+M^2-2\omega V-V^2\right)^{-1}
\left(\omega^2+\overleftarrow{\partial}_x\cdot\overrightarrow{\partial}_x
+M^2-V^2\right)\Big]\,,
\label{eq:vac2}
\end{equation}
The functional trace ${\rm Tr}^\prime$ no longer includes the integral over 
the time variable. Note that the vacuum energy $E_{\rm vac}$ still contains the 
contribution from the free case ($A_\mu\equiv0$), while the  Casimir energy 
$E_{\rm cas} = E_{\rm vac}-E_{\rm vac}^{(0)}$ is the \emph{change} in the vacuum 
energy due to the photon background. We can now integrate the spatial
derivatives by parts and assume that the boundary terms vanish either
because the space interval is infinite and the fields drop off sufficiently 
fast, or because the Boson fields obey periodic boundary conditions in a finite 
space interval. We obtain
\begin{eqnarray}
E_{\rm vac}&=&
=i\int \frac{d\omega}{2\pi}\,\omega\,{\rm Tr}^\prime\, 
\Big[\left(-\omega^2-\partial_x^2+M^2-2\omega V-V^2\right)^{-1}
\frac{\partial}{\partial \omega}
\left(-\omega^2-\partial_x^2+m^2-2\omega V-V^2\right)\Big] +c_1 \cr\cr
&=&i\int \frac{d\omega}{2\pi}\,\omega\, \frac{\partial}{\partial \omega}
{\rm Tr}^\prime\, {\rm log} \Big[-\omega^2-\partial_x^2+M^2-2\omega V-V^2\Big]+c_2
\label{eq:vac3}
\end{eqnarray}
where the constants $c_1$ and $c_2$ are independent of the background potential $V$
(though they may require regularization). Comparison with 
Eq.~(\ref{eq:action}) immediately shows that 
\begin{equation}
E_{\rm vac}=-\frac{1}{T}S[A_\mu]\,,
\label{eq:vac4}
\end{equation}
where the (infinitely large) time interval $T$ entered by restoring
the full functional trace. The last equation is of course well known
\cite{Weinberg:1995mt}, but still a bit surprising in the
present context, since  the Hamiltonian derived from
Eq.~(\ref{eq:lag}) still contains time derivatives and the problem is
thus not completely static, even though $V(x)$ depends only on
the space variable. The additional time derivatives in the interaction
complicate the Legendre transformation between energy and action per
unit time, but Eq.~(\ref{eq:vac4}) still holds in this case. 

The canonical approach is also unconventional but straightforward; it can 
be formulated as an application of the random phase approximation \cite{Ri80}. 
After Fourier transforming the time-dependence, the field equations turn
into wave-equations for bound states $\phi_n(x)$ with discrete energies 
$|\omega_n| < M$, and for scattering solutions $\phi_k^{(\pm)}(x)$ with 
energies\footnote{The notation for the energy eigenvalues is that 
subscripts $m$ and $n$ denote discrete bound states.}
$\pm\omega_k$ (where $\omega_k=\sqrt{k^2 + M^2} > 0$),
\begin{eqnarray}
\Big[ - \partial_x^2 - V(x)^2 + 2 \omega_n\,V(x) + M^2 \Big]\, \phi_n(x) 
&=& \omega_n^2\,\phi_n(x) 
\label{eq:bs1}\\[2mm]
\Big[- \partial_x^2 - V(x)^2 \pm 2 \omega_k\,V(x) + M^2 \Big]\,\phi_k^{(\pm)}(x)
&=& \omega_k^2\,\phi_k^{(\pm)}(x)\,.
\label{eq:con1}
\end{eqnarray}
Since the theory is not CP-invariant, the existence of a bound state with 
energy $\omega_n$ does not imply the existence of a corresponding bound state
with negative energy $(-\omega_n)$, and likewise $\phi_k^{(+)}(x)\ne\phi_k^{(-)}(x)$
for the scattering states. 

To determine the normalization of these states, let us first consider the bound state
equation (\ref{eq:bs1}). Multiplying this equation by $\phi_m(x)$ from the right,
integrating over space and subtracting the same equation with the replacement 
$m\leftrightarrow n$ immediately shows
\begin{equation}
(\omega_m-\omega_n)\int dx\, \phi_m(x)\left[
\omega_m + \omega_n - 2 V(x) \right]\,\phi_n(x)\,=0\,.
\label{eq:ortho0}
\end{equation}
The norm of the bound state wave functions can thus be determined by the orthogonality 
relation\footnote{The sign on the right hand side of Eq.~(\ref{eq:ortho0}) follows for 
$V=0$ because the bound state energies are real. For $V \neq 0$, we do not consider 
potentials that are so attractive that they change the sign of a bound state energy 
that started off at $\pm M$ for $V(x)\equiv0$. Such potentials would 
lead to a charged ground state.}
\begin{equation}
\int dx\, \phi_m(x)\big[\omega_m+\omega_n-2V(x)\big]\phi_n(x)
=\delta_{mn}\, {\rm sgn}(\omega_m)\,.
\label{eq:ortho1}
\end{equation}
A similar consideration for the scattering states leads to
\begin{equation}
\int dx\, \phi_k^{(s)}(x)\big[s\omega_k+r\omega_{q}
-2V(x)\big]\phi_{q}^{(r)}(x)=2\pi s \delta_{s r}\,\delta(k-q)\,.
\label{eq:ortho2}
\end{equation}
Of course, bound state and continuum solutions are orthogonal
\begin{equation}
\int dx\, \phi_k^{(s)}(x)\big[s\omega_k+\omega_n
-2V(x)\big]\phi_n(x)=0\,.
\label{eq:ortho3}
\end{equation}
With these orthonormality conditions (and the corresponding completeness relations, 
{\it cf}.~Eq.~(\ref{eq:comp1}) below), we can decompose the field operator in the 
usual fashion,
\begin{equation}
\Phi(x,t)=\sum_{\omega_n\ge 0}\,a_n \,\phi_n(x) {\rm e}^{-i\omega_n t}
+\sum_{\omega_n<0} b_n^\dagger\phi_n(x) {\rm e}^{-i\omega_n t}
+\int_0^\infty \frac{dk}{2\pi}\left[
a(k)\phi_k^{(+)}(x){\rm e}^{-i\omega_k t}
+b^\dagger(k)\phi_k^{(-)}(x){\rm e}^{i\omega_k t}\right]\,.
\label{eq:fcomp}
\end{equation}
To quantize the model, we have to impose the canonical equal time commutation 
relations between the field $\Phi$ and its conjugate momentum $\Pi$, which involves 
a covariant rather than an ordinary time derivative,
\begin{equation}
\Pi(x,t)=\frac{\partial \mathcal{L}}{\partial \dot{\Phi}(x,t)}
=\dot{\Phi}^\ast(x,t)-iV(x)\Phi^\ast(x,t)\,.
\label{eq:can1}
\end{equation}
Inserting the decomposition Eq.~(\ref{eq:fcomp}) in these commutation relations, and 
using the orthogonality conditions, Eqs.~(\ref{eq:ortho1})--(\ref{eq:ortho3}) to extract 
the expansion coefficients, we obtain the usual ladder algebra
\begin{alignat}{2}
\left[a_n,a_m^\dagger\right] &=\delta_{nm}\,, &
\left[b_n,b_m^\dagger\right] &=\delta_{nm}
\nonumber
\\[2mm]
\left[a(k),a^\dagger(q)\right]&= 2\pi\delta(k-q)\,,&
\qquad\qquad\left[b(k),b^\dagger(q)\right] & =2\pi\delta(k-q)
\label{eq:ladder1}
\end{alignat}
where all other combinations vanish. We use these creation and annihilation 
operators to construct a Fock space and compute the vacuum matrix element of the 
energy density operator
\begin{equation}
2T^{00}(x)=
\dot{\Phi}^\ast\dot{\Phi}^\dagger
-\Phi^\ast\left(\partial_t^2+2iV\partial_t\right)\Phi^\dagger
+\dot{\Phi}\dot{\Phi}^\dagger
-\Phi\left(\partial_t^2-2iV\partial_t\right)\Phi^\dagger+\ldots\,,
\label{eq:T00a}
\end{equation}
where the ellipsis refers to total space derivatives that do not contribute 
to the integrated energy. We obtain by textbook techniques
\begin{equation}
\big\langle \,0\, \big|\,T^{00}(x)\,\big|\,0\,\big\rangle =
\sum_n \omega_n\, \phi_n(x)\big[\omega_n-V(x)\big]\phi_n(x)
+\sum_{s=\pm} \int_0^\infty \frac{dk}{2\pi}\,\omega_k\,
\phi_k^{(s)}(x)\,\big[\omega_k-s\,V(x)\big]\,\phi_k^{(s)}(x) +\ldots\,.
\label{eq:T00b}
\end{equation}
At this point, we see that the vacuum energy density indeed contains  a 
non-standard factor which involves the interaction explicitly. However, 
the same factor appears in Eqs.~(\ref{eq:ortho1}) and (\ref{eq:ortho2}). 
When this is taken into account, the total vacuum energy becomes
\begin{equation}
E_{\rm vac} =\int dx\,\big \langle \,0\, \big|\,T^{00}(x)\,\big|\,0
\,\big\rangle= \frac{1}{2}\sum_n |\omega_n|+\frac{L}{2}\int_0^\infty
\frac{dk}{2\pi}\, \left(2\omega_k\right)\,,
\label{eq:casimir}
\end{equation}
where $L = 2 \pi \delta(0)$ is the spatial volume (length) of the 
universe\footnote{The factor of two under the integral comes from the sum over 
positive and negative frequencies.}. This is precisely the usual sum over 
zero point energies. Of course, this expression is highly divergent and only 
formal at this point. (We will eventually have to subtract the
non--interacting case $V=0$, and add appropriate counterterms in higher
dimensions.) The important conclusion is, however, that the frequency
dependence in the interaction does not alter the expression for the vacuum 
energy in the present case.

Let us next study how the spectral method applies in this model. We first note 
the completeness relations for the scattering and bound states that are 
compatible with Eqs.~(\ref{eq:ortho1})
and (\ref{eq:ortho2}),
\begin{alignat}{3}
\sum_n|\omega_n|\,\phi_n(x)\,\phi_n(y) &+ \int_0^\infty 
\frac{dk}{2\pi}\,\sum_{s=\pm}\,\omega_k\,\phi_k^{(s)}(x)\,\phi_k^{(s)}(y) 
&\,\,=\,\,& \delta(x-y)
\nonumber 
\\[2mm]
\sum_n\mathrm{sgn}(\omega_n)\,\phi_n(x)\,\phi_n(y) &+ \int_0^\infty 
\frac{dk}{2\pi}\,\sum_{s=\pm}\,s\, \phi_k^{(s)}(x)\,\phi_k^{(s)}(y) 
&\,\,=\,\,& 0\,.
\label{eq:comp1}
\end{alignat}
The second relation would be a trivial identity in a CP--invariant model, 
but here it follows as a consistency condition on Eqs.~(\ref{eq:ortho1}) 
and {(\ref{eq:ortho2})}.  We separate the Green's function in frequency 
space ($|\omega|\ge M$) in three pieces, {\it viz.} the bound state part 
plus the negative and positive frequency contributions from the continuum
\begin{align}
G(x,y,\omega)&= G^{\rm (b.s.)}(x,y,\omega)
+\Theta(\omega)G^{(+)}(x,y,k)+\Theta(-\omega)G^{(-)}(x,y,k)
\nonumber
\\[2mm]
G^{\rm (b.s.)}(x,y,\omega)&=\sum_n {\rm sgn}(\omega_n)
(\omega_n+\omega) \frac{\phi_n(x)\phi_n(y)}
{\omega_n^2-\omega^2}
\nonumber\\[2mm]
G^{(\pm)}(x,y,k)&=\sum_{s=\pm}\int_0^\infty \frac{dq}{2\pi}
\left(\omega_q\pm s\omega_k\right)
\frac{\phi_q^{(s)}(x)\phi_q^{(s)}(y)}{q^2-(k+i\epsilon)^2}\,,
\label{eq:green1}
\end{align}
where $k \equiv \sqrt{\omega^2-M^2} > 0$. Because of the completeness 
relations (\ref{eq:comp1}), this Green's function obeys
\begin{equation}
\left[\omega^2-2\omega V(x)+\partial_x^2+V^2(x)-M^2\right]
G(x,y,\omega)=-\delta(x-y)\,.
\label{eq:green2}
\end{equation}
The boundary conditions on the Green's function (the $i\epsilon$ prescription) 
were chosen such that the poles for $\mathsf{Im}(k)>0$ are those of the bound 
state wave numbers, {\it i.e.}~$G$ has the same analytic structure as in ordinary 
scattering theory, {\it cf.}~Eq.~(\ref{eq:int2}).

Since the square of the bound state wave functions is real in our normalization, 
only the continuum contribution to the Green's function can have an imaginary part,
which stems from  $\mathsf{Im}[q^2-(k+i\epsilon)^2]^{-1}=\pi\delta(q^2-k^2)$.
It is then straightforward to show from Eq.~(\ref{eq:T00b}) that the vacuum 
energy density is
\begin{equation}
u(x)=
\sum_n\omega_n \phi_n(x)\big[\omega_n-V(x)\big]\phi_n(x)
+\frac{1}{\pi}\int_{-\infty}^\infty d\omega\, \omega\big[\omega-V(x)\big]\,
\mathsf{Im}\left[G(x,x,\omega)-G^{(0)}(x,x,\omega)\right]\,.
\label{eq:green3}
\end{equation}
This is the analog of Eq.~(\ref{eq:int6}) in the standard spectral method.
It should be stressed again that the modified normalization of states leads 
to non-standard coefficients that involve the background potential explicitly.

The last step is to relate the vacuum energy to scattering data. To do so,
we find the analog of Eq.~(\ref{eq:int3}) from the
wave--equation~(\ref{eq:bs1})
\begin{equation}
W\left[f^{(s)}_k(R),\phi^{(s)}_q(R)\right]=F^{(s)}(k)+\int_0^R dx
\left[\omega_k^2-\omega_q^2
-2s(\omega_k-\omega_q)V(x)\right] f^{(s)}_k(x)\phi^{(s)}_q(x)\,,
\label{eq:wred1}
\end{equation}
where the superscript $s=\pm1$ on the scattering wave functions denotes the 
positive and negative frequency channels. Retracing the steps that 
in the standard approach led to Eq.~(\ref{eq:int5}), we find, 
\begin{equation}
0=\sum_{s=\pm}\left\{
\frac{\partial}{\partial k}\,{\rm ln}\left|F^{(s)}(k)\right|
-i\frac{\partial}{\partial k} \delta^{(s)}(k)\right\}
+2\,\frac{k}{\omega_k}\int_0^\infty dx\, \left[\omega_k-V(x)\right]
\left[G(x,x,\omega_k)-G^{(0)}(x,x,\omega_k)\right]\,,
\label{eq:wred2}
\end{equation}
after summation over both frequency channels. Taking the imaginary part
and comparing with the spatial integral of Eq.~(\ref{eq:green3}), we obtain 
\begin{equation}
E_{\rm vac}=\int_0^\infty dx\, u(x)=\frac{1}{2}\sum_n\,|\omega_n|
+\int_0^\infty \frac{dk}{2\pi}\,\omega_k\frac{\partial}{\partial k}\delta(k)\,,
\label{eq:edfinal}
\end{equation}
where $\delta(k)=\delta^{(+)}(k)+\delta^{(-)}(k)$ is the sum of scattering 
phase shifts for positive and negative frequencies at a given momentum $k$.
Formally, Eq.~(\ref{eq:edfinal}) equals the earlier expression Eq.~(\ref{eq:int7}) 
obtained in the case of frequency independent interactions. No extra terms emerge 
because the non--trivial terms that involve the background potential are the same 
in the energy density, Eq.~(\ref{eq:T00b}) as in the frequency derivative of 
the wave--equation, Eq.~(\ref{eq:wred2}).

To summarize, we have now seen in three approaches that the frequency dependent 
interactions in scalar electrodynamics do not lead to modifications in the standard 
expressions, {\it cf.} Eqs.~(\ref{eq:vac4}),~(\ref{eq:casimir}) and~(\ref{eq:edfinal}).
The energy density, Green's functions, and scattering data all showed non--trivial 
alterations involving the background potential $V(x)$ explicitly, yet in the final 
result all these unconventional contributions canceled. Thus, the standard spectral 
approach is versatile enough to handle frequency dependent interactions without extra 
terms, at least in the case of scalar electrodynamics. A similar structure can also be 
found in the bound state approach to the three flavor Skyrme model 
\cite{Callan:1985hy,Callan:1987xt,Blaizot:1988ge,Scoccola:1998eq}
wherein a linear frequency dependence arises from the Wess--Zumino term. It is 
therefore expected that the spectral method in this case also receives
no extra terms (though renormalizability is an issue in the model).

However, the derivations above have also shown that the necessary cancellations are 
quite subtle, and it is by no means obvious whether this finding extends to 
\emph{all} frequency dependent interactions which often occur in effective models
when heavier fields are integrated out. In the case of scalar
electrodynamics we have been in the fortunate position that we were in
control of the wave equations and the energy (density) at any
stage. In effective models this need not be true anymore.  In the next sections, we will therefore study two prototype cases, {\it viz.} the effective 
action induced by the decoupling of some heavy degree of freedom, and
electrodynamics in a dielectric. 

\section{Heavy Meson}

In this section, we consider a model of two scalar boson fields that are
coupled through the static but space--dependent background $V(\vek{x})$,
\begin{align}
 \mathcal{L} = \frac{1}{2}\,\partial_\mu \Phi\,\partial^\mu \Phi - \frac{M^2}{2}\,\Phi^2 
 + \frac{1}{2}\,\partial_\mu \varphi\,\partial^\mu \varphi - \frac{m^2}{2}\,\varphi^2 + 
 V(\vek{x})\,\Phi\,\varphi\,.
 \label{201}
\end{align}
We take the field $\Phi$ much heavier, $M \gg m$, and study the effective model 
for the light boson $\varphi$ that arises when $\Phi$ is integrated out. This is a typical 
situation that occurs in a wide range of applications, from solid state to particle 
physics. At long distances, $\varphi$ is expected to be the dominant degree of freedom, 
but the resulting effective action need not be local in space or time. Thus, we have again 
a model where we can study an effective, frequency dependent interaction and 
compare with a full theory that is local and renormalizable, {\it i.e.}~where the exact 
vacuum energy can be computed reliably. We will also keep the number $d$ of space 
dimensions $d$ arbitrary, although we will eventually take $d=1$ in a numerical 
example to dispense with renormalization issues. 

Let us study the effective $\varphi$-model first. Since the light field $\varphi$ acts as 
a source for the heavy field, integrating out~$\Phi$ yields the effective action
\begin{align}
S_{\rm eff}[\varphi] = \frac{1}{2}\int d^{d+1}x\,\big[ \partial_\mu  \varphi\,\partial^\mu 
\varphi - m^2\,\varphi^2 \big] - \frac{1}{2}\,\int d^{d+1}x \int d^{d+1}y\,
\varphi(y)\,V(\vek{y})\int \frac{d^{d+1}k}{(2\pi)^{d+1}}\,\frac{e^{i k \cdot (x-y)}}
{k^2 - M^2 + i \epsilon}\,V(\vek{x})\,\varphi(x)\,,
\end{align}
where a field-independent term involving $\text{det}(\partial^2+M^2)$ has been 
dropped. Since the coupling $V(\vek{x})$ is time-independent, a Fourier transformation of the 
light field,
\begin{align}
\varphi(x) = \int \frac{d\omega}{2\pi}\,\varphi(\omega,\vek{x})\,e^{- i \omega\,x_0}
\label{201x}
\end{align}
yields the effective model
\begin{align}
S_{\rm eff}[\varphi] = \frac{1}{2}\,\int d^{d+1} x\,\big[\partial_\mu \varphi\,\partial^\mu \varphi - 
m^2 \,\varphi^2\big]
+ \frac{1}{2}\,\int d^{d}x \int d^{d}y\int\frac{d\omega}{2\pi}\,\varphi(\omega,\vek{y})\,
\KK(\omega,\vek{x},\vek{y})\,\varphi(\omega,\vek{x})^\ast\,,
\label{202}
\end{align}
where the interaction kernel is non--local is space,
\begin{align}
 \KK(\omega,\vek{x},\vek{y}) = - V(\vek{y})\,\int
\frac{d^dk}{(2\pi)^d}\,
\frac{{\rm e}^{-i\vek{k}\cdot (\vek{x}-\vek{y})}}
{\omega^2-\vek{k}^2-M^2+i\epsilon}\,V(\vek{x})\,.
 \label{204}
\end{align}
We are interested in an approximation at large space distances $|\vek{x}-\vek{y}| \gg 1$,
where the integral in Eq.~(\ref{204}) is dominated by small $|\vek{k}|$. Neglecting the 
$\vek{k}^2$ in the denominator compared to $M^2$ yields the {\it local approximation}
\begin{align}
\KK (\omega,\vek{x},\vek{y}) \approx \delta^{(d)}(\vek{x}-\vek{y})\,
\frac{- V(\vek{x})^2}{\omega^2 - M^2}\,.
\label{205}
\end{align}
In view of the initial covariance of the model, it is of course tempting to also neglect the 
frequency dependence $\omega^2$ in comparison to the large mass $M$ in the denominator of 
Eq.~(\ref{205}). However, this is not adequate in the present case because~\textbf{(i)} 
we are interested in the effective model at large \emph{spatial} distances but arbitrary 
frequencies, and~\textbf{(ii)} the omission of~$\omega^2$ in Eq.~(\ref{205}) in general 
leads to a loss of renormalizability.

Let us next look at the energy density. From the full model (\ref{201}), we obtain the field 
equations
\begin{equation}
(\partial^2 + M^2)\,\Phi = V(\vek{x})\,\varphi
\qquad {\rm and}\qquad
(\partial^2+ m^2)\,\varphi = V(\vek{x})\,\Phi\,,
\label{210}
\end{equation}
and the energy density
\begin{align}
T^{00} = \frac{1}{2}\,\Big( \dot{\Phi}^2 - \Phi\,\ddot{\Phi} + \dot{\varphi}^2- 
\varphi\,\ddot{\varphi} \Big) + \text{total space derivatives}\,,
\label{212}
\end{align}
where we have used the field equations to produce the total derivative terms. Next we 
introduce the frequency by Fourier transforming both fields as in Eq.~(\ref{201x}), which 
yields field equations for each frequency $\omega$ separately,
\begin{align}
- \nabla^2\,\Phi_\omega(\vek{x}) &= (\omega^2 - M^2)\,\Phi_\omega(\vek{x}) + V(\vek{x})\,
\varphi_\omega(\vek{x})\,,
\nonumber
\\[2mm]
- \nabla^2 \,\varphi_\omega(\vek{x}) &= (\omega^2 - m^2)\,\varphi_\omega(\vek{x}) +  
V(\vek{x})\,\Phi_\omega(\vek{x})\,.
\label{208}
\end{align}
The local approximation (\ref{205}) amounts to $|\vek{k}|^2 \ll M^2$, {\it i.e.}~the 
spatial derivatives of $\Phi$ in the first equation can be neglected compared to the 
right hand side. As a consequence, $\Phi_\omega \approx V\,\varphi_\omega/(M^2-\omega^2)$,
and the second field equation  turns into the effective equation of motion
\begin{align}
- \nabla^2\,\varphi_\omega \approx (\omega^2 - m^2)\,\varphi_\omega - \frac{V(\vek{x})^2}{
\omega^2 - M^2 + i \epsilon}\,\varphi_\omega\,,
\label{220}
\end{align}
where we have reinstated the Feynman pole prescription as in Eq.~(\ref{204}). From this 
wave--equation we derive an orthogonality relation via a calculation similar to the one 
yielding Eqs.~(\ref{eq:ortho1})--(\ref{eq:ortho3}),
\begin{equation}
\int d^dx\, \varphi_{\omega}(x)\left[1+\frac{V(\vek{x})^2}
{\left(\omega^2-M^2+i\epsilon\right)\left(\omega^{\prime2}-M^2+i\epsilon\right)}
\right]\varphi_{\omega^\prime}(x)=0
\qquad {\rm for} \qquad \omega\ne\omega^\prime\,.
\label{eq:220a}
\end{equation}
In the same way, we can use the local approximation for $\Phi_\omega$ directly
in the energy density Eq.~(\ref{212}), which in view of Eq.~(\ref{eq:220a}) 
becomes a single integral over frequency modes
\begin{align}
T^{00}(\vek{x}) = \int \frac{d\omega}{2\pi}\,T^{00}_\omega(\vek{x}) \equiv \int 
\frac{d\omega}{2\pi}\,\omega^2\left[1+\frac{V(\vek{x})^2}{(\omega^2-M^2+i\epsilon)^2}
\right]\,\varphi_\omega(\vek{x})^2\,,
\label{219}
\end{align}
where we have omitted contributions that vanish upon spatial integration.
We stress that in coordinate space where $\omega\to i\partial_t$, the total energy
$\int d^d x\, T^{00}(\vek{x})$ is conserved by the wave equation~(\ref{220}) of the 
local approximation alone. For $\omega=\omega^\prime$ the 
factors in square brackets in Eqs.~(\ref{eq:220a}) and~(\ref{219}) agree.
This reflects a consistency condition discussed in the appendix.

Let us now translate Eq.~(\ref{219}) into a phase shift formula. We start with the 
time--independent field equation (\ref{220}) at a fixed frequency $\omega$ and now work 
in $d=1$ space dimension 
\begin{align}
\Big[- \partial_x^2 + U_\omega(x)\Big] \,\varphi_\omega(x) = (\omega^2-m^2)\,\varphi_\omega(x)
\qquad {\rm with} \qquad
U_\omega(x) = \frac{V(x)^2}{\omega^2 - M^2 + i \epsilon}\,.
\end{align}
We can now proceed as in section II and introduce Jost, regular and 
scattering solutions to this equation. The Wronskian Eq.~(\ref{eq:int3}) becomes
\begin{align}
W\big[ \,f_k(R), \phi_q(R)\,\big] = F(k) + \int_0^R dx\,\left[ k^2 - q^2 - U_{\omega_k}(x) 
+ U_{\omega_q}(x) \right] \,f_k(x)\,\phi_q(x)\,,
\label{eq:wryam}
\end{align}
where $f$ and $\phi$ are the Jost and regular solution, respectively. 
Proceeding as in section II we obtain, instead of Eq.~(\ref{eq:int5}),
\begin{align}
\frac{\partial}{\partial k}\,\ln F(k) &= - \int_0^\infty dx\,\left \{ 2 k \,\left(
\frac{f_k(x)\,\phi_k(x)}{F(k)} - f_k^{(0)}(x)\,\phi_k^{(0)}(x)\right) - 
\left[ \frac{\partial U_\omega(x)}{\partial k}\right]\,\frac{f_k(x)\,\phi_k(x)}{F(k)}\right \}
\nonumber \\[2mm]
&= -2 k \int_0^\infty dx\,\left\{ \left[1 + \frac{V(x)^2}{(\omega^2 - M^2 + i \epsilon)^2}
\right]\, G(x,x;k) - G^{(0)}(x,x;k) \right\}\,,
\label{230}
\end{align}
where the Green's function is defined as in Eq.~(\ref{eq:int2}). The additional term arises 
because the potential $U_\omega(x)$ is $k$-dependent. However, the modifying factor in front 
of the Green's function is exactly what is needed in the energy density Eq.~(\ref{219}).
In other words, the standard phase shift formula (\ref{eq:int7}) with the phase shift expressed 
by the Jost function corresponds, in view of Eq.~(\ref{230}), to the energy density
\begin{align}
u(x) = \int \frac{dk}{\pi}\,\sqrt{k^2 + m^2}\,k \,\mathsf{Im}\,\left( 
\left[1 + \frac{V(x)^2}{(\omega^2 - M^2 + i \epsilon)^2}\right]
G(x,x,k) - G^{(0)}(x,x,k)\right) + \cdots\,.
\label{240}
\end{align}
Retracing the steps in Ref.~\cite{Graham:2002xq}, it is found that the modified energy 
density Eq.~(\ref{240}) agrees with the modified expression, Eq.~(\ref{219}) obtained from the 
full field equations in the local approximation. As a result, the vacuum energy 
$E_{\rm vac}$ of our frequency dependent interaction in the local approximation takes the 
standard form (\ref{eq:int7}) despite the fact that unusual coefficients appear in the 
course of the derivation. Though we have not rigorously derived the wave--equation and 
the energy density from the full theory, we were able to apply equivalent approximations 
and thus obtain an energy density that is consistent with the wave--equation in the
effective model.

\begin{figure}[ht]
 \centering
 \includegraphics[width=11cm,height=5cm,keepaspectratio=false]{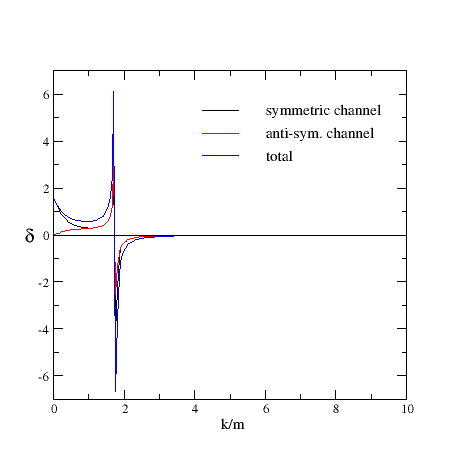}
 \caption{Phase shifts for $\epsilon=10^{-10}$, $a=2$, $c=1$, and $M/m=2$; 
the threshold pole is at $k_T/m = \sqrt{3}$.}
 \label{fig:1}
\end{figure}

\medskip\noindent
To close this section, let us finally check the validity of the local
approximation, Eq.~(\ref{205}),
by explicit numerical calculations. Let us begin with the effective model. As pointed out above,
we can follow the standard approach laid out in section II, with the 
potential replaced by $U_\omega(x)$. Further following Ref.~\cite{Graham:2002xq}, we can 
parameterize the Jost solution at energies $\omega > m$ through the ansatz 
$f_k(x)={\rm e}^{ikx+i\beta(k,x)}$ and, from the field equation~(\ref{220}), obtain a 
non--linear differential equation for the complex function $\beta(k,x)$,
\begin{align}
 -i \beta''(k,x) + 2 k \beta'(k,x) + \big[ \beta'(k,x)\big]^2 + 
\frac{V(x)^2}{k^2 + m^2 - M^2 + i \epsilon} = 0\,,
\label{250}
\end{align}
where the prime denotes the spatial derivative. The solution to this differential
equation is subject to the boundary conditions $\beta(k,\infty)=\beta'(k,\infty)=0$. The 
vacuum energy can now be computed from the standard spectral formula, provided that the phase 
shifts are defined through the Jost function in the conventional way. We consider symmetric 
potentials $V(x)=V(-x)$ and decompose the scattering problem on the real axis into two 
scattering problems on the half--axis $x\ge0$. The phase shifts in the anti- and symmetric 
channels are respectively given by
\begin{align}
\delta_-(k) &=-\frac{1}{2}\,\mathsf{Re}\,\lim_{x\to 0}\,\Big[\beta(k,x)-\beta(-k,x)\big]
\nonumber \\[2mm]
\delta_+(k) &=\delta_-(k)+\frac{1}{2}\,\mathsf{Im}\,\lim_{x\to 0}\,\ln \frac{
\beta'(-k,x)-k}{\beta'(k,x)+k}\,. 
\label{252}
\end{align}
The combinations of Eq.~(\ref{250}) and (\ref{252}) allows us to compute the total phase shift 
$\delta(k)=\delta_+(k)+\delta_-(k)$, 
which then yields the vacuum energy from Eq.~(\ref{eq:int7}), after integration by parts using 
Levinson's theorem \cite{Graham:2002xq},
\begin{align}
E_{\rm vac} = - \int_0^\infty  \frac{dk}{2\pi}\,\frac{k}{\sqrt{k^2 + m^2}}\,
\delta(k) + \frac{1}{2}\sum_{n} \big(|\omega_n| - m\big)\,.
\label{254}
\end{align}
Eqs.~(\ref{250})--(\ref{254}) are a complete system to compute the vacuum energy $E_{\rm vac}$ 
in the effective model using the local approximation. The only numerical issue arises due to 
the pole in Eq.~(\ref{250}) at threshold, $k_T=\sqrt{M^2-m^2}$. Our field theoretic derivation in 
Minkowski space has provided the correct $i\epsilon$-prescription to circumvent this pole, but 
it is still possible that the direct limit $\epsilon \searrow 0$ is numerically infeasible.
To settle this issue, we have taken a sample background profile
\begin{align}
V(x) = m\,c \, e^{-(mx/a)^2}\,, 
\label{282}
\end{align}
with numerical constants $a$ and $b$. Table \ref{tab:1} shows results obtained
for $c=1$, $a=2$ with masses $M/m = 2$ and the threshold at $k_T/m = \sqrt{3}$. The integrals 
in the table correspond to the relevant integral in Eq.~(\ref{254}) --- up to a factor of 
$1/(2\pi)$ ---  but now taken in a finite range covering the threshold, {\it i.e.} $I_L$ 
covers the range $k \in [3m/2, k_T]$, while $I_R$ refers to $[k_T, 2m]$. We then 
take $\epsilon = 10^{-n}$ and increase $n$ to let $\epsilon \to 0$. As can be clearly 
seen, both integrals \emph{saturate} around $n=8$, {\it i.e.} the limit $\epsilon\to 0$ 
in Eq.~(\ref{250}) can be taken literally. Figure \ref{fig:1} shows the resulting 
phase shifts for a value $\epsilon=10^{-10}$, where the numerics have converged.
\renewcommand{\arraystretch}{1.4} 
\begin{table}[t]
\centering
\begin{tabular}{c|ccccccccc}
\toprule
$n$ & 2 & 3 & 4 & 5 & 6 & 7 & 8 & 9 & 10 \\ \colrule
$\,\,\,I_L\,\,\,$ & 0.499 &  0.605 &  0.660 & 0.687 & 0.700 & 0.706 & 0.709 & 0.710 & 0.711 \\
$I_R$ & -0.642 & -0.652 & -0.653 &-0.653 &-0.653 &-0.653 &-0.653 &-0.653 &-0.653 \\
\botrule
\end{tabular}
\caption{\label{tab:1}The left and right integrals explained in the main text, 
as functions of the position of the pole, $\epsilon=10^{-n}$. 
The model parameters are as in figure~\ref{fig:1}.}
\end{table}
\renewcommand{\arraystretch}{1.0}

Next, we turn to the vacuum energy in the full model, Eq.~(\ref{201}). This is a simple 
\emph{coupled channel problem} which is discussed at length in the literature 
\cite{Chadan:1977pq, Newton:1982qc}. The idea is to write down the field equations 
for the vector $(\varphi,\Phi)$ and then put two of the four linear independent 
vector solutions (labeled by subscripts 1,2) in the columns of a 
$(2 \times 2)$ matrix $\mathcal{F}$, 
\begin{equation}
\begin{pmatrix} \varphi_1 & \varphi_2 \cr
 \Phi_1 & \Phi_2\end{pmatrix}=\mathcal{F}\cdot
\begin{pmatrix}{\rm e}^{ikx} & 0 \cr
0 & {\rm e}^{iKx} \end{pmatrix} \,.
\label{eq:defF}
\end{equation}
The other two solutions are obtained by complex conjugation 
because the original wave--equation~(\ref{208}) is real. The matrix $\mathcal{F}$
is suitable for a scattering problem because it parameterizes out--going plane 
waves with the boundary conditions $\mathcal{F}=1$ and $\mathcal{F}^\prime=0$ at
spatial infinity. For a fixed energy $\omega$,
it obeys the differential equation
\begin{align}
\mathcal{F}''(x) = \begin{pmatrix}0 & V(x) \\ V(x) & 0\end{pmatrix} \cdot\mathcal{F}(x)
-2i \mathcal{F}'(x)\cdot\begin{pmatrix} k & 0 \\ 0 & K \end{pmatrix}
+ \left[ \mathcal{F}(x)\,,\,
\begin{pmatrix} k^2 & 0 \\ 0 & K^2 \end{pmatrix} \right]
\label{292}
\end{align}
where $k = \sqrt{\omega^2 - m^2}$ and $K = \sqrt{\omega^2 - M^2}$ above threshold,
and $K = i \sqrt{M^2 - \omega^2}$ below threshold.  The latter case produces an exponential 
decay for $\Phi$ at large distances, and thus ensures that no flux occurs in this
channel below threshold. 
In the antisymmetric channel, we find the $S$-matrix and phase shift, respectively, 
from the Jost solution
to Eq.~(\ref{292})
\begin{align}
\mathcal{S}_- = \lim_{x \to 0}\mathcal{F}^{-1}\cdot \mathcal{F}^\ast
\qquad{\rm and}\qquad
\delta_- = \frac{-i}{2}\,\ln \mathrm{det}\,\mathcal{S}\,. 
\label{280}
\end{align}
In principle, we should compute the phase shift below threshold only from the open channel,
$\delta_{-}= (-i/2)\,\ln\left(\mathcal{S}_{-}\right)_{11}$. Eq.~(\ref{280}) still works in 
this regime, because $\left(\mathcal{S}_{-}\right)_{21} = 0$ and 
$\left(\mathcal{S}_{-}\right)_{22}=1$.

In the symmetric channel, the situation is slightly more complicated. From the 
requirement that the derivative of the wave--function vanishes at the origin, we find
\begin{align}
\mathcal{S}_+ = - \lim_{x \to 0}\,\left[ \mathcal{F}' + \mathcal{F}\,
\begin{pmatrix} ik & 0 \\ 0 & i K \end{pmatrix}\right]^{-1} \cdot 
\left[ {\mathcal{F}^\ast}' - \mathcal{F}^\ast\,\begin{pmatrix} ik & 0 \\ 0 & i K 
\end{pmatrix}\right]\,.
\end{align}
Again, $\delta_+ = (-i/2)\,\ln\det\mathcal S_+$ above threshold, and 
$\delta_+ = (-i/2)\,\ln \left(\mathcal{S}_{+}\right)_{11}$ below. This time, however, 
$\left(\mathcal{S}_{+}\right)_{22}=-1$ below threshold and we have to add an extra 
constant in $\delta_+ = - \pi/2 - i/2\,\ln\det \mathcal{S}_+$.\footnote{This constant 
is also necessary to satisfy Levinson's theorem in the symmetric 
channel: Without coupling between the two fields but with some
self--interactions their phase shifts at zero momentum would be an odd multiple 
of $\pi/2$. Adiabatically switching on the coupling maintains that value below 
threshold.}

\begin{figure}[ht!]
\centering
\includegraphics[width=6cm,height=5cm,keepaspectratio=false]{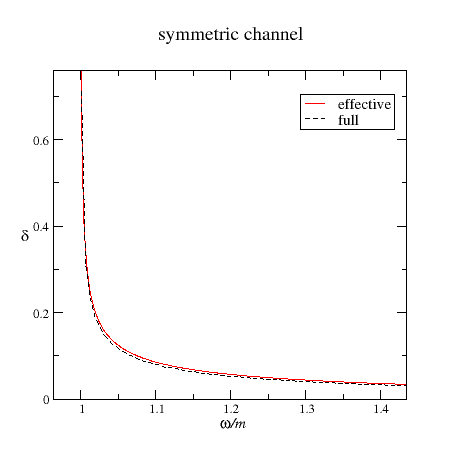}
\hspace*{2cm}
\includegraphics[width=6cm,height=5cm,keepaspectratio=false]{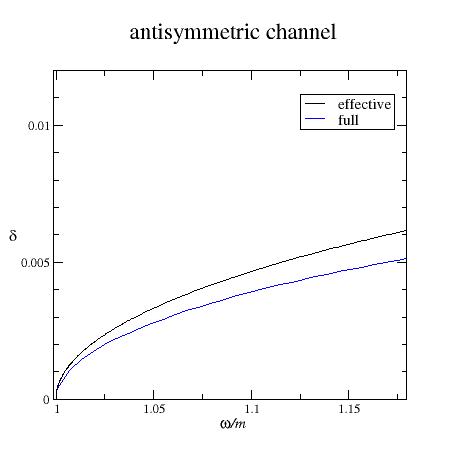}
\caption{\label{fig:2}
Comparison of the phase shifts in full and effective models for low energies. 
In the symmetric channel $\delta\to\pi/2$ as $\omega\to m$,
suggesting a single bound state. The model parameters are $M/m=4$, $a=2$
and $c=1$.}
\end{figure}

Figure \ref{fig:2} compares the low energy phase shifts in the full and 
effective model for the background potential Eq.~(\ref{282}) with $M/m=4$.
As can be seen, the local approximation works fairly well at low
energies, even if $M$ is not dramatically larger than $m$.

Finally, we compare the vacuum energies of the full and the effective model, 
which we split into continuum and bound state contributions according to 
Eqs.~(\ref{eq:int7}) and (\ref{254}). The bound states can be computed by 
a variety of methods such as \emph{shooting}, direct \emph{diagonalization}  
of the Schr\"odinger operator in a suitable function basis, or from the roots
of the Jost function at imaginary momentum $k=i\kappa$. In all methods we find, 
besides the usual bound states, formal solutions with $\omega^2 < 0$ as well. 
These solutions represent unphysical states with an amplitude increasing 
exponentially in time, {\it i.e.} they are undamped \emph{resonances} that 
must not be included in the vacuum energy, although they are required to 
satisfy Levinson's theorem, which merely counts the number of roots of 
the Jost function in $k$--space. 

In table \ref{tab:2}, we have listed the vacuum energy (in units of $m$) for 
our preferred background Eq.~(\ref{282}) and $a=2$, $c=6$. As can be seen, 
there is still a considerable difference between the full and exact model,
especially for small mass ratios $M/m$. The discrepancy is almost completely 
due to the continuum contribution $E_\delta$, whose relative error decreases
significantly as $M/m$ gets larger. Unfortunately, the numerics become very 
delicate for larger masses, because the exponential damping ${\rm e}^{-|K|x}$ 
becomes very small and cannot be accurately compared to data of order one.

\renewcommand{\arraystretch}{1.4} 
\begin{table}[t!]
\centering
\begin{tabular}{c c c c c }
\toprule
$\quad M/m\quad$ & $\quad E_\delta\quad$ & $E_{\text{bs}}^2 (\text{sym})$ & 
$E_{\text{bs}}^2 (\text{asym})$ & $\quad E_{\text{vac}}\quad $ \\ \colrule
2  &-0.20 & -2.58 ; 0.77 &-0.60 & -0.26 \\
3  & 0.29 & -1.22        & 0.42 & 0.11 \\
4  & 0.36 & -0.29        & 0.88 & 0.33 \\
5  & 0.32 & 0.23         & 0.99 & 0.06 \\
6  & 0.27 & 0.52         & $0.99^\ast$  & 0.13 \\
7  & 0.22 & 0.68         & $0.99^\ast$  & 0.13 \\
\botrule
\end{tabular}
\hspace*{13mm}
\begin{tabular}{c c c c c }
\toprule
$\quad M/m\quad$ & $\quad E_\delta\quad$ & $E_{\text{bs}}^2 (\text{sym})$ & 
$E_{\text{bs}}^2 (\text{asym})$ & $\quad E_{\text{vac}}\quad $ \\ \colrule
2  & 0.75 & -2.83 ;-0.08 &-1.27; 0.77 & 0.69 \\
3  & 0.80 & -1.34        & 0.18 & 0.51 \\
4  & 0.65 & -0.34        & 0.83 & 0.60 \\
5  & 0.52 & 0.21         & $0.99^\ast$     & 0.25 \\
6  & 0.41 & 0.51         & $0.99^\ast$     & 0.26 \\
7  & 0.33 & 0.68         & $0.99^\ast$     & 0.24 \\
\botrule
\end{tabular}
\caption{\label{tab:2}Contributions to the vacuum energy for the full 
model (\emph{left}) and the effective model (\emph{right}) at various mass 
ratios $M/m$. Parameters for the background coupling are $a=2$, $c=6$ and 
all energies are given in units of the light meson mass~$m$. The phase shift 
contribution $E_\delta$ refers to the \emph{binding energy}, 
Eq.~(\ref{254}). Bound state energies are given as absolute values.
The vacuum energy is given by  
$E_{\rm vac}=E_\delta+\frac{1}{2}\,\sum_n\big(|E_{\rm bs}|-m\big)$\,,
where the sum only includes true bound states with $E_{\rm bs}^2 > 0$. 
Asterisks mark bound states that have not been found numerically since they are too 
close to threshold. They must exist because of Levinson's theorem and they must 
have binding energies of less than $0.01$ since our numerical algorithm would have 
detected them otherwise. Numerically, their contribution to the vacuum energy is 
hence negligible.}
\end{table}
\renewcommand{\arraystretch}{1.0}

\section{Dielectric}

The introduction of a permeability, $\mu$ and a permittivity, $\epsilon$ to describe the 
interaction of electromagnetic fields with matter is a prime example for an effective 
model. It arises from integrating out the microscopic matter degrees of freedom of the 
fundamental QED formulation and leads to the Maxwell equations in matter
\begin{equation}
\nabla\cdot\vek{D}=4\pi\rho\,,\qquad\qquad
\nabla\times\vek{E}+\partial_t \vek{B}=0\,,\qquad\qquad
\nabla\cdot\vek{B}=0\qquad {\rm and}\qquad
\nabla\times\vek{H}-\partial_t \vek{D}=4\pi\vek{j}\,,
\label{eq:maxmatter}
\end{equation}
where 
$\vek{D}(t,\vek{x})=\int dt^\prime
\epsilon(t^\prime,\vek{x})\,\vek{E}(t-t^\prime,\vek{x})$ and
$\vek{B}(t,\vek{x})=\int dt^\prime
\mu(t^\prime,\vek{x})\,\vek{H}(t-t^\prime,\vek{x})$ and
information about the charge carriers in the material is contained in
the phenomenological quantities $\mu$ and $\epsilon$.  For simplicity
we restrict our attention to the case of $\mu = 1$.

We are interested in vacuum polarization energies so there are no
external charges or  currents and we set $\rho=0$ and $\vek{j\,}=0$.
Frequency dependence arises in this model in two ways:  first,
via the explicit time derivatives in Eq.~(\ref{eq:maxmatter}), which
generate factors of frequency multiplying the interaction, and second 
via the frequency dependence of the dielectric function $\epsilon$ itself.  
The latter arises because the response of any realistic material
must go to zero for large frequency.  Furthermore, by the
Kramers--Kronig relations, enforcing causality requires that a
frequency-dependent dielectric function have a nontrivial
imaginary part, representing dissipation.

Of course, electromagnetic Casimir fluctuations do not lead to energy
dissipation in a dielectric material; they exist in equilibrium with
fluctuations in the microscopic matter degrees of freedom. Such an effective 
model can be described by a partition function in quantum statistical 
mechanics, but not by a canonical quantum field theory, since it lacks a 
strictly conserved energy. In the statistical mechanics approach 
\cite{Rahi:2009hm,Kenneth06,Kenneth:2007jk}, 
the partition function is defined on the imaginary frequency axis, where 
the dielectric function is strictly real (even with dissipation). Because the 
nonlocality in time introduced by the frequency dependence in $\epsilon$ always 
takes the form of a convolution, one can factorize the problem according to 
frequency. To compute the partition function, one then integrates over all field
configurations; this integral is dominated by contributions arising from the 
saddle-point integration around each solution to the classical equation of motion 
for a fixed frequency. From the partition function one obtains the free energy, 
which at $T=0$ gives the Casimir energy as a statistical ensemble average. This 
is the approach that has been adopted in
Refs. \cite{Rahi:2009hm,Kenneth06,Kenneth:2007jk}.
By relating the resulting functional integral to the free energy 
one finds an energy density given by the integral 
over space of $\frac{1}{2}(\vek{E}\cdot\vek{D}+\vek{B}^2)$. 
The saddle point approximation to this functional integral
gives the fields as solutions to Maxwell's equations in matter. This
statistical mechanics picture for the energy is
consistent with the classical thermodynamics result that fluctuations
of the charge density due to exposure to an electric field contribute
the spatial integral of $\frac{1}{2}\vek{E}\cdot\vek{D}$ to the free
energy \cite{Panofsky,Jackson}.  The problem is stationary and it is
suitable to introduce Fourier modes\footnote{Since the 
photon is massless, frequency $\omega$  and wave--number $k$ are synonymous.}
$\vek{E}_k(\vek{x})=\int dt\, {\rm e}^{ikt}\, \vek{E}(\vek{x},t)$ and
$\vek{B}_k(\vek{x})=\int dt\, {\rm e}^{ikt}\, \vek{B}(\vek{x},t)$.
Their contributions to the energy are then summed over frequency
$k$. The frequency dependence in the interaction 
then leads to additional terms 
in the relation~(\ref{eq:int5}) between the integrated wave--function 
and the Jost function for imaginary frequencies\footnote{The statistical 
mechanics picture imposes imaginary frequencies from the outset.}, 
as described in Ref.~\cite{Graham:2013yza}.  In this case, we obtain the
Casimir energy of a single Fourier mode from the integral 
over space of
\begin{equation}
u_{I}(\vek{x})=\frac{1}{2}\epsilon_k
\vek{E}_k^{2}+\frac{1}{2}\vek{B}_k^{2} \,.
\label{eq:dielectric1}
\end{equation}

Here, we will follow an alternative path at the cost of violation of
the Kramers--Kronig relations.  We will stay within quantum field
theory to consider a conserved energy that is derived from Maxwell's 
equations in matter~(\ref{eq:maxmatter}). As outlined in the appendix, it 
is an awkward expression in coordinate space. However, when expressed in 
terms of the Fourier modes, it takes a compact from, that {\it e.g.} has 
been used in Refs.~\cite{Milton:2010yw,PhysRevA.81.033812,PhysRevA.84.053813},
see Eq.~(\ref{eq:dielectric2}) below. We will find that in this case, the 
additional terms in the relationship between the Jost function and the 
integrated wave--functions cancel with the additional terms in the energy 
expression, leaving the same expressions as arise for a frequency--independent 
potential. In this case, we can carry out a derivation within
field theory and find the strictly conserved energy
to be the integral over space of
\begin{equation}
u_{II}(\vek{x})=\frac{1}{2}\frac{\partial[k\epsilon_k]}{\partial
k}\,\vek{E}_k^{2}
+\frac{1}{2}\vek{B}_k^{2} \,.
\label{eq:dielectric2}
\end{equation}
Both Eqs.~(\ref{eq:dielectric1}) and (\ref{eq:dielectric2}) give the
contribution of a single mode.
Formally the time-independence of $\int d^3x\, u_{II}(\vek{x})$
can be verified for all cases in which $\epsilon_k$ has a Taylor 
expansion in~$k^2$, but not~$k$. This condition is, however, not fulfilled by 
any physically motivated dielectric functions such as Eq.~(\ref{eq:dielfct}) 
below.  In particular, this sort of Taylor expansion is not compatible with 
the Kramers--Kronig relations for spectral functions. Nevertheless this
case allows us to investigate the effect of frequency dependent
interactions in the context of spectral methods applied to vacuum energies.

Assuming a spherically symmetric dielectric  
$\epsilon_k(r)=\int dt\, {\rm e}^{ikt}\,\epsilon(t,\vek{x})$ leads to 
the so--called Mie--model \cite{Mie:1908,Newton:1982qc}.  With these
preliminaries Maxwell's equations~(\ref{eq:maxmatter}) can be
transformed into decoupled  second order differential equations for
scalar fields $\varphi_k(\vek{x})$ and $\phi_k(\vek{x})$ by the ans\"atze
\begin{align}
&{\rm TE:}\qquad
&\vek{E}_k(\vek{x})= k\, \nabla \times [\varphi_k(\vek{x})\, \vek{x}] 
\qquad {\rm and} \qquad 
&\vek{B}_k(\vek{x})=i \nabla \times \left(\nabla \times 
\left[\varphi_k(\vek{x})\, \vek{x}\right]\right) \cr\cr
&{\rm TM:}\qquad
&\vek{B}_k(\vek{x})= -ik \, \nabla \times 
\left[\phi_k(\vek{x})\, \vek{x}\right]
\qquad {\rm and} \qquad 
&\vek{E}_k(\vek{x})=\frac{-1}{\epsilon_k(r)} \nabla \times \left(\nabla
\times \left[\phi_k(\vek{x})\, \vek{x}\right]\right)\,,
\hspace{3cm}
\label{eq:goodbadE}
\end{align}
for the transverse electric (TE) and transverse magnetic (TM) modes, respectively.
The spherical decompositions~(with angular momentum channels $\ell=1,2,\ldots$)
\begin{equation}
\varphi_{k,\ell m}(\vek{x}) = \frac{1}{\sqrt{\ell(\ell+1)}}
Y_{\ell}^m(\theta,\phi) \frac{1}{r}f_{k,\ell}(r) 
\qquad {\rm and} \qquad
\phi_{k,\ell m}(\vek{x}) = 
\frac{\sqrt{\epsilon_k(r)}}{\sqrt{\ell(\ell+1)}} Y_{\ell}^m(\theta,\phi) 
\frac{1}{r} g_{k,\ell}(r)
\label{eq:spherical1}
\end{equation}
allow to consistently\footnote{The factor $\sqrt{\epsilon_k(r)}$ is required to obtain
an Hermitian scattering operator in the case that $\epsilon_k(r)$ is real.} compute 
scattering data for a fixed frequency $k$ from the differential equations
\begin{eqnarray}
k^2 \epsilon_k(r) f_{k,\ell}(r) &=&
-f_{k,\ell}''(r) + \frac{\ell(\ell+1)}{r^2}f_{k,\ell}(r)
\qquad {\rm and} \cr\cr
k^2\epsilon_k(r) g_{k,\ell}(r)&=&-g_{k,\ell}''(r) +\frac{\ell(\ell+1)}{r^2}g_{k,\ell}(r) +
\left[\frac{3 \epsilon_k^{\prime2}(r)}{4 \epsilon_k^2(r)}
- \frac{\epsilon_k^{\prime\prime}(r)}{2 \epsilon_k(r)} \right] g_{k,\ell}(r)\,,
\label{eq:spherical2}
\end{eqnarray}
where primes denote derivatives with respect to the radial coordinate. We have also 
elevated the permittivity to a dielectric function of the frequency $k$ and indicated 
that by a subscript. This frequency dependence is unavoidable to maintain the renormalizability 
which should be {preserved} because it is manifest in the underlying quantum electrodynamics. 
From scattering theory we know that a non--trivial metric, as represented by the left hand 
sides of Eqs.~(\ref{eq:spherical2}), leads to linearly rising phase shifts at large $k$. 
This increase cannot be compensated by any finite number of Born subtractions. In the 
language of Feynman diagrams it corresponds to a quadratic derivative interaction with the 
background. Multiplying such a vertex with a propagator approaches a constant in the 
ultra--violet producing infinitely many divergent diagrams. Hence, to maintain 
renormalizability, the deviation of the dielectric function from unity must vanish like 
$1/k^2$ at large frequencies. Note that this reasoning is similar as for the 
effective model considered in section IV. The dielectric function is commonly
parameterized in terms of the imaginary frequency~$\kappa=-ik$,
\begin{equation}
\epsilon_{i\kappa}(r)=1+\frac{p(r)}{\kappa^2\lambda^2+\frac{\pi}{\sigma_P}\kappa}\,,
\label{eq:dielfct}
\end{equation}
where $\lambda$ is the so--called plasma wave--length whose inverse corresponds to a 
material cut--off in the ultra--violet according to the above discussion. Furthermore, $\sigma_P$
is the conductivity whose main purpose in the context of Casimir energy calculations is
to circumvent infra--red divergences. If the spherically symmetric profile function $p(r)$ 
is strongly peaked at a specified radius, the above is a well--suited effective model for 
the interaction of photons with a dielectric sphere.

Angular integration yields the radial energy densities (omitting the angular momentum
label for simplicity)
\begin{eqnarray}
2u_{I}^{\rm(TE)}(k,r)&=&\epsilon_k(r)k^2 f_k^2(r)+[(rf_k(r))^\prime]^2+\ell(\ell+1)f_k^2(r)\cr\cr
2u_{I}^{\rm(TM)}(k,r)&=&
g_k^{\prime2}(r)+\frac{\epsilon_k^\prime(r)}{\epsilon_k(r)}g_k(r)g_k^\prime(r)
+\frac{\epsilon^{\prime2}_k(r)}{4\epsilon_k^2(r)}g_k^2(r)
+\frac{\ell(\ell+1)}{r^2}g_k^2(r)+k^2\epsilon_k(r)g_k^2(r)
\label{eq:edensI1}
\end{eqnarray}
and
\begin{eqnarray}
2u_{II}^{\rm(TE)}(k,r)&=&k^2\left[\epsilon_k(r)+k\dot{\epsilon}_k(r)\right]f_k^2(r) 
+[(rf_k(r))^\prime]^2 +\ell(\ell+1)f_k^2(r)\cr\cr
2u_{II}^{\rm(TM)}(k,r)&=&\left[1+k\frac{\dot{\epsilon}_k(r)}{\epsilon_k(r)}\right]
\left[g_k^{\prime2}(r)+\frac{\epsilon^\prime_k(r)}{\epsilon_k(r)}g_k(r)g_k^\prime(r)
+\frac{\epsilon^{\prime2}_k(r)}{4\epsilon^2_k(r)}g_k^2(r)
+\frac{\ell(\ell+1)}{r^2}g_k^2(r)\right]+k^2\epsilon_k(r)g_k^2(r)
\label{eq:edensII1}
\end{eqnarray}
for the two candidates of Eqs.~(\ref{eq:dielectric1}) and
(\ref{eq:dielectric2}). In contrast to the previous 
sections the dot now denotes a derivative with respect to frequency, 
$\dot{\epsilon}_k=\partial \epsilon_k/\partial k$. Using the differential 
equations~(\ref{eq:spherical2}) and omitting total space derivative terms simplifies 
the densities considerably in case ($I$) \cite{Graham:2013yza}
\begin{equation}
u_{I}^{\rm(TE)}(k,r)=\epsilon_k(r)k^2 f_k^2(r) 
\qquad {\rm and} \qquad
u_{I}^{\rm(TM)}(k,r)=\epsilon_k(r)k^2 g_k^2(r)\,,
\label{eq:edensI2}
\end{equation}
but not in case ($II$)
\begin{eqnarray}
u_{II}^{\rm(TE)}(k,r)&=&k^2\left[\epsilon_k(r)+\frac{k}{2}\dot{\epsilon}_k(r)\right] f_k^2(r)
\cr\cr
u_{II}^{\rm(TM)}(k,r)&=&
k^2\left[\epsilon_k(r)+\frac{k}{2}\dot{\epsilon}_k(r)\right]g_k^2(r)
+\frac{k}{4}\left[
\frac{\dot{\epsilon}^{\prime\prime}_k(r)}{\epsilon_k(r)}
-3\frac{\dot{\epsilon}^\prime_k(r){\epsilon}^\prime_k(r)}{\epsilon^2_k(r)}
-\frac{\dot{\epsilon}_k(r){\epsilon}^{\prime\prime}_k(r)}{\epsilon^2_k(r)}
+3\frac{\dot{\epsilon}_k(r){\epsilon}^{\prime2}_k(r)}{\epsilon^3_k(r)}
\right]g_k^2(r)\,.
\label{eq:edensII2}
\end{eqnarray}

In the next step we want to express the energy density by scattering data. As 
previously, {\it cf.} Eqs.~(\ref{eq:int3}),~(\ref{eq:wred1}) and~(\ref{eq:wryam}),
we start by computing the Wronskian between the regular and Jost solutions at
different frequencies. This involves the field equations~(\ref{eq:spherical2}) from
which we identify
\begin{equation}
U_k^{\rm (TE)}(r)=k^2\left[1-\epsilon_k(r)\right]
\qquad {\rm and} \qquad
U_k^{\rm (TM)}(r)=k^2\left[1-\epsilon_k(r)\right]
+\left[\frac{3 \epsilon_k^{\prime2}(r)}{4 \epsilon_k^2(r)}
- \frac{\epsilon_k^{\prime\prime}(r)}{2 \epsilon_k(r)} \right]
\label{eq:wrdiel1}
\end{equation}
to be substituted into, say, Eq.~(\ref{eq:wryam}). It remains to compare the frequency
derivatives
\begin{eqnarray}
-\frac{\partial}{\partial k}\left[k^2+U_k^{\rm (TE)}(r)\right]
&=&2k\epsilon_k(r)+k^2\dot{\epsilon}_k(r) 
\qquad {\rm and} \qquad \cr\cr
-\frac{\partial}{\partial k}\left[k^2+U_k^{\rm (TM)}(r)\right]
&=&2k\epsilon_k(r)+k^2\dot{\epsilon}_k(r)
-\frac{3\dot{\epsilon}^\prime_k(r)\epsilon^\prime_k(r)}{2\epsilon^2_k(r)}
+\frac{3\dot{\epsilon}_k(r)\epsilon^{\prime2}_k(r)}{2\epsilon^3_k(r)}
+\frac{\dot{\epsilon}^{\prime\prime}_k(r)}{2\epsilon_k(r)}
-\frac{\dot{\epsilon}_k(r)\epsilon^{\prime\prime}_k(r)}{2\epsilon^2_k(r)}
\label{eq:wrdiel2}
\end{eqnarray}
with the coefficient functions in the expressions for energy densities.
Obviously there are significant differences in case~($I$) but up to the
additional factor $k/2$ it agrees with the formula
for the energy density in case~($II$). This implies that the latter
case will have the standard phase  shift expression for the vacuum
polarization energy, Eq.~(\ref{eq:int1})  while case ($I$) has
additional contributions. These contributions have been  studied in
Ref.~\cite{Graham:2013yza} for a dielectric sphere and were found  to
yield an attractive force.

To summarize this section, we have shown that in the absence of dissipation, 
we have a formulation that is rigorous and well founded in quantum field
theory, while in the presence of dissipation such a full quantum field 
theory formulation is not available and we need to resort to a statistical 
mechanics approach. In the latter case one employs the conventional expression 
for the energy with a constant dielectric (type ($I$)), which leads to
extra terms  in our spectral analysis, while the unconventional form
($II$), that one  obtains in the former case, does not.

\section{Conclusions}

We have discussed several models in which quantum fluctuations interact
with a (static) background through frequency dependent couplings. In such 
scenarios the conjugate field momenta, and thus all observables computed as 
Noether charges, involve the interaction explicitly, rather than only 
intrinsically via the solutions to the wave--equation. Furthermore it is 
not obvious that energy and action functionals of even static configurations 
are proportional. We have therefore studied the question whether
the simple phase shift formula, Eq.~(\ref{eq:master}), for the 
vacuum polarization energy remains valid in such cases.  On top of the 
arguments above, the validity of this formula is not at all evident since 
the major ingredients for its derivation in the frequency independent 
case no longer hold: On the (canonical) quantum field theory side 
we must re--examine the relationship between the energy density and the 
Green's function, while on the scattering problem side we must re--consider 
the relationship between the derivative of the phase shift and the
appropriate spatial integral over the Green's function. Modifications
may arise because this relationship is deduced from the derivative of 
the wave--equation with respect to frequency.

In an effective model the frequency dependence is commonly non--polynomial
, which prevents us from applying canonical quantization. As a result, 
statements about the vacuum expectation values of the energy can only
be made by comparison with  classical analogs.

In the simplest of all possible models with frequency dependent interactions, 
scalar electrodynamics, the phase shift formula continues to hold because the 
modifications on the field theory and scattering sides compensate each other. 
In effective models which emerge from integrating out fundamental degrees of 
freedom, approximations that ignore the frequency dependence usually spoil
renormalizability and thus must be avoided in the context of quantum
energies. As seen in the simple model of section IV, the resulting
frequency dependence in the effective theory is non--polynomial. This
makes it difficult to construct a conserved energy functional 
from the wave--equations from within the effective model. Generally, 
such a construction is possible only when the wave--equation is a polynomial
of the frequency squared, {\it cf.}~appendix \ref{app:a}. 

In the model we considered, we were fortunate enough to be able to trace the 
energy functional from the fundamental theory. When we applied the \emph{same}
approximation to the energy density as for the wave--equation when
integrating out the heavier field, we observed that the phase shift
formula also holds in such a scenario. This result comes about because the 
local approximation to the energy density gives the same result as the 
polynomial construction principle applied to the wave--equation in the local 
approximation. After obtaining this result formally in the effective model 
we have also verified it numerically by comparing to the full theory in the 
regime where the adopted approximations are valid.

As a third example we have considered the physically interesting problem of a dielectric. 
The relevant wave--equations are Maxwell's equation in matter. Assuming spherical 
symmetry for the background leads to the Mie model. This formulation represents an 
effective model because the fundamental interactions between the charge carriers 
and quantum fluctuations of the electromagnetic field are imitated phenomenologically.
We have compared two formulations of this effective theory. When we include dissipation, 
as required by the Kramers--Kronig relations, we do not have a quantum field theory 
formulation of the problem. Instead a statistical mechanics ensemble 
average \cite{Rahi:2009hm,Kenneth06,Kenneth:2007jk} leads to an energy density of type ($I$) with 
the ensuing extra terms in the phase shift formula. They can then play an important 
role in the calculation of Casimir self--stresses \cite{Graham:2013yza}.
Without dissipation, a Taylor expansion in the frequency squared is permitted 
and a conserved quantity can be constructed from the wave--equation, leading to 
an energy density of type ($II$), which does not lead to extra terms in the phase 
shift analysis.  For forces between rigid bodies the situation is simpler, since 
one can consider the stress tensor $T_{\mu\nu}$ outside the dielectric, where
$T_{\mu\nu}$ is uniquely defined and the fields are obtained
from the Mie equation.

\bigskip

\begin{acknowledgments}
The authors acknowledge communication with G.\ Barton and K.\ A.\ Milton that
motivated this research. N.\ G.\ thanks G.\ Bimonte, T.\ Emig, R.\
L.\ Jaffe, M.\ Kardar, and M.\ Kr\"uger for helpful conversations.
N.\ G.\ was supported in part by the National Science Foundation (NSF)
through grant PHY-1213456.  H.\ W.\ was supported by the National
Research Foundation (NRF), Grant No. 77454.
\end{acknowledgments}

\appendix

\section{Energy from Wave--Equation}
\label{app:a}

In this appendix we explain the statement of section V concerning the derivation 
of an energy functional when the wave--equation has a Taylor expansion in the 
frequency squared. This analysis is based on analogies with classical systems.

Let $\phi_k(\vek{x})$ be a solution to the wave--equation with frequency\footnote{We 
continue to denote the frequency by $k$ even though we do not require the field to 
be massless in this appendix.} $k$. We assume that the wave--equation only contains
even powers of $k$
\begin{equation}
\left(k^2+\alpha_2 k^4+\alpha_3 k^6+\ldots\right)\phi_k=\hat{O}\phi_k\,.
\label{eq:b1}
\end{equation}
Here $\alpha_i$ may be spatial functions and $\hat{O}$ is some Hermitian
operator that contains a static potential and perhaps some spatial derivatives.
Let us consider a term like $k^{2n}\phi_k(\vek{x})$. In coordinate space it 
amounts to $(-1)^n\partial_t^{2n}\phi(t,\vek{x})$. We choose the overall sign 
such that the standard $n=1$ term has a positive coefficient, multiply this 
term by $\partial_t\phi(t,\vek{x})$ and identify a total derivative: 
\begin{eqnarray}
(-1)^{n+1}\left(\partial_t\phi\right)\left(\partial_t^{2n}\phi\right)&=&(-1)^{n+1}\left\{
\partial_t\left[\left(\partial_t\phi\right)\left(\partial_t^{2n-1}\phi\right)\right]
-\left(\partial_t^2\phi\right)\left(\partial_t^{2n-1}\phi\right)\right\}\cr\cr
&=&(-1)^{n+1}\left\{
\partial_t\left[\left(\partial_t\phi\right)\left(\partial_t^{2n-1}\phi\right)\right]
-\partial_t\left[\left(\partial_t^2\phi\right)\left(\partial_t^{2n-2}\phi\right)\right]
+\left(\partial_t^3\phi\right)\left(\partial_t^{2n-2}\phi\right)\right\}\cr\cr
&\vdots& \cr\cr
&=&(-1)^n\left\{\partial_t\sum_{j=1}^{n-1}(-1)^j
\left(\partial_t^j\phi\right)\left(\partial_t^{2n-j}\phi\right)
+(-1)^n\left(\partial_t^n\phi\right)\left(\partial_t^{n+1}\phi\right)\right\}\cr\cr
&=&\partial_t\left\{\sum_{j=1}^{n-1}(-1)^{j+n}
\left(\partial_t^j\phi\right)\left(\partial_t^{2n-j}\phi\right)
+\frac{1}{2}\left(\partial_t^n\phi\right)\left(\partial_t^n\phi\right)\right\}\,.
\label{eq:a1}
\end{eqnarray}
Hence we identify the contribution to the density of the conserved energy as
\begin{equation}
u_n(t,\vek{x})=\sum_{j=1}^{n-1}(-1)^{j+n}
\left(\partial_t^j\phi\right)\left(\partial_t^{2n-j}\phi\right)
+\frac{1}{2}\left(\partial_t^n\phi\right)\left(\partial_t^n\phi\right)\,.
\label{eq:a2}
\end{equation}
In frequency space the energy density is a double integral over $k$ and $k^\prime$,
$u_n(t,\vek{x})=\int \frac{dk}{2\pi} \int \frac{dk^\prime}{2\pi}
\phi_{k^\prime}\left[\ldots\right]\phi_{k}{\rm e}^{i(k+k^\prime)t}$,
where the ellipsis replaces the frequency representation of the operators in 
Eq.~(\ref{eq:a2}). Then the vacuum matrix element projects this bilocal object onto 
its diagonal ($k=-k^\prime$) terms in a similar way, since they contribute solely to 
the total energy $\int d^3x \sum_n u_n(t,\vek{x})$; that calculation is provided 
below (classically one can argue that energy conservation 
allows one to average over time, which also implements this projection). 
The diagonal projection becomes time-independent,
\begin{eqnarray}
\tilde{u}_n(k)&=&(-1)^n\phi_{-k}\left\{\sum_{j=1}^{n-1}(-1)^j
\left(ik\right)^j\left(-ik\right)^{2n-j}
+\frac{1}{2}(-1)^n\left(ik\right)^n\left(-ik\right)^n
\right\}\phi_k\cr\cr
&=&(-1)^n\phi_{-k}\left\{\sum_{j=1}^{n-1}
(i)^{2n}k^{2n}+\frac{1}{2}(-1)^nk^{2n}\right\}\phi_k
=\frac{1}{2}\phi_{-k}\left(2n-1\right)k^{2n}\phi_k
=\frac{1}{2}\phi_{-k}\left\{k^2\partial_k\left(k^{2n-1}\right)\right\}\phi_k\,,
\label{eq:a3}
\end{eqnarray}
where the last equation relates to the type ($II$) energy of section V and we 
have not written the frequency integral explicitly. To show that this construction 
does not work for odd powers of the frequency we merely need to consider 
\begin{equation}
\left(\partial_t\phi\right)\left(\partial_t^3\phi\right)=
\partial_t\left[\left(\partial_t\phi\right)\left(\partial_t^2\phi\right)\right]
-\left(\partial_t^2\phi\right)\left(\partial_t^2\phi\right)\,.
\label{eq:a4}
\end{equation}
The second term on the right hand side cannot be expressed as a total derivative
and further application of the product rule merely yields a trivial identity. In the 
case of scalar electrodynamics (section II) a complex quantum field was studied. 
Models that couple odd powers of the frequency via different field components are 
permissible. It is also obvious that the construction does not straightforwardly apply 
to a non--polynomial dependence on frequency either, because the product rule cannot
be used. Such a dependence, however, is essential for a renormalizable model. 

The frequency independent term on the right hand side of Eq.~(\ref{eq:b1})
contributes $\frac{1}{2}\int d^dx\,\phi_{-k}\hat{O}\phi_k$ to the energy. 
This contribution can be expressed via the wave--equation~(\ref{eq:b1}) and 
adds $(1/2)\phi_{-k}\,\alpha_n k^{2n}\,\phi_k$ to $\tilde{u}_n$, 
{\it cf.}~Eq.~(\ref{212}). This changes the coefficient in Eq.~(\ref{eq:a3}) to
\begin{equation}
\frac{1}{2}(2n-1)k^{2n}+\frac{1}{2}k^{2n}
=k^2\frac{\partial}{\partial k^2}\, k^{2n}\,.
\label{eq:b3}
\end{equation}
Said another way, once the potential type contribution to the energy has been 
eliminated via the wave--equations, the energy factor is given by 
$k^2\frac{\partial}{\partial k^2}\,V_k(\vek{x})$, where $V_k(\vek{x})$
represents the complete frequency dependence in the wave--equation. In this way
Eqs.~(\ref{220}) and~(\ref{219}) are consistent. The application of this 
prescription to Eq.~(\ref{eq:spherical2}) yields Eq.~(\ref{eq:edensII1}). We also 
note that this result is kindred to the frequency derivatives that appear in the 
derivation of the phase shift formula after Eqs.~(\ref{eq:int3}),~(\ref{eq:wred1}) 
and~(\ref{230})
since $k^2\frac{\partial}{\partial k^2}\,V_k(\vek{x})=
\frac{k}{2}\frac{\partial}{\partial k}\,V_k(\vek{x})$.

In the context of Eq.~(\ref{eq:a3}) we have noted that the energy density is bilocal 
in frequency but only the diagonal projection is relevant for the total energy. This 
projection arises from the orthogonality condition between solutions with different 
frequencies. Again assuming that $V_k(\vek{x})$ is even in $k$, this condition reads
\begin{equation}
\int d^{d}x f_{k^\prime}(\vek{x})
\frac{V_k(\vek{x})-V_{k^\prime}(\vek{x})}{k^2-k^{\prime2}}f_k(\vek{x})
\propto \delta(k^2-k^{\prime2})\,.
\label{eq:c1}
\end{equation}
From the geometrical series identity
\begin{equation}
\frac{k^{2n}-k^{\prime2n}}{k^2-k^{\prime 2}}(k-k^\prime)^2=
2\sum_{j=1}^{n-1}\left[(-k^\prime)^jk^{2n-j}+(-k)^jk^{\prime2n-j}\right]
+2(-kk^\prime)^n+k^{2n}+k^{\prime2n}
\label{eq:c2}
\end{equation}
one can verify order by order that the total energy constructed above indeed turns 
into a single frequency integral and is time independent. The first three terms on 
the right hand side of Eq.~(\ref{eq:c2}) arise from Eq.~(\ref{eq:a2}) after 
identifying $k$ and $k^\prime$. The last two stem from eliminating the explicit 
$\hat{O}$ contribution via the wave--equations. Furthermore, note that the metric 
factor for the normalization of the wave--function extracted from Eq.~(\ref{eq:c1})
in the limit $k^\prime\to k$ leads to the construction prescription for constructing 
the weight factor in the energy.

The above discussion was carried out for a simple scalar field. However, it is 
straightforwardly applied to electromagnetism in matter starting from
the wave--equations 
\begin{equation}
\nabla\times\vek{E}+\partial_t \vek{B}=0
\qquad{\rm and}\qquad
\nabla\times\vek{B}
-\partial_t\left[\epsilon_{-i\partial_t}\vek{E}\right]=0
\label{eq:a5}
\end{equation}
in coordinate space. Subtraction of the two equations after scalar multiplication of
the first with $\vek{B}$ and the second with $\vek{E}$ yields a left hand side 
\begin{equation}
\nabla\cdot\left(\vek{E}\times\vek{B}\right)=
\vek{B}\cdot\partial_t\vek{B}
+\vek{E}\cdot\partial_t\,\epsilon_{-i\partial_t}\vek{E}
\label{eq:a6}
\end{equation}
which is the divergence of the Poynting vector and vanishes after spatial 
integration. If the dielectric function had the expansion 
$\epsilon_{-i\partial_t}=1+\sum_{n=1}^N \alpha_n \partial_t^{2n}$
the manipulations analogous to Eq.~(\ref{eq:a1}) would
then suggest that the coordinate representation $\int d^3x\,
u_{II}(t,\vek{x})$ is constant in time. Note that in Eq.~(\ref{eq:a6}) the 
potential type term $\phi\hat{O}\phi\sim\vek{B}^2$ has not yet been eliminated by the
wave--equation.  Thus we indeed require $\frac{\partial}{\partial
k}(k\epsilon_k)$ and not $\frac{\partial}{\partial
k^2}(k^2\epsilon_k)$ as the coefficient of~$\vek{E}^2$.

As for the scalar case, the generalization of this construction to odd powers in 
the frequency requires mixing of components. This can be accomplished by a tensor 
relation\footnote{We note that this is an argument solely within field theory and
that this tensor should not be confused with the physical permittivity, which is 
always symmetric. It can, however, by viewed as modeling optical activity 
\cite{Tsao:1993}.} between the displacement $\vek{D}$ and the electric field 
$\vek{E}$. In particular odd powers of the frequency must be accompanied by 
an anti--symmetric tensor. The feature of such an anti--symmetric internal 
structure is the disappearance of the left over term in Eq.~(\ref{eq:a4}). The 
simple example 
\begin{equation}
\vek{D}=\vek{E}+\Vek{a}\times\dot{\vek{E}}\,.
\label{eq:a7}
\end{equation}
where $\Vek{a}$ is some vector with dimension of time that may vary in space, 
illustrates this point. To construct a conserved quantity we again find the contribution to the 
divergence of the Poynting vector
\begin{equation}
\vek{E}\cdot\partial_t \vek{D}
=\vek{E}\cdot\partial_t \left[\vek{E}+\Vek{a}\times\dot{\vek{E}}\right]
=\partial_t\left[\frac{1}{2}\vek{E}^2+\vek{E}\cdot(\Vek{a}\times\dot{\vek{E}})\right]\,.
\label{eq:a8}
\end{equation}
The object in square brackets is the contribution to the energy density. 
In frequency space it has the diagonal projection
\begin{equation}
\frac{1}{2}(E_{-k})_i\left[
\delta_{ij}-2ik\epsilon_{ilj}a_l\right](E_k)_j\,,
\label{eq:a9}
\end{equation}
which again has the kernel $\frac{\partial}{\partial k}(k\epsilon_k)$
of $u_{II}$. Standard textbook \cite{Landau,Schwinger} derivations of
this kernel consider electric fields that are a superposition of waves
with frequencies in a small bandwidth. Obviously the 
arguments presented in this appendix do not rely on this simplifying
assumption.


\begin{thebibliography}{38}
\expandafter\ifx\csname natexlab\endcsname\relax\def\natexlab#1{#1}\fi
\expandafter\ifx\csname bibnamefont\endcsname\relax
  \def\bibnamefont#1{#1}\fi
\expandafter\ifx\csname bibfnamefont\endcsname\relax
  \def\bibfnamefont#1{#1}\fi
\expandafter\ifx\csname citenamefont\endcsname\relax
  \def\citenamefont#1{#1}\fi
\expandafter\ifx\csname url\endcsname\relax
  \def\url#1{\texttt{#1}}\fi
\expandafter\ifx\csname urlprefix\endcsname\relax\def\urlprefix{URL }\fi
\providecommand{\bibinfo}[2]{#2}
\providecommand{\eprint}[2][]{\url{#2}}

\bibitem[{\citenamefont{Graham et~al.}(2009)\citenamefont{Graham, Quandt, and
  Weigel}}]{Graham:2009zz}
\bibinfo{author}{\bibfnamefont{N.}~\bibnamefont{Graham}},
  \bibinfo{author}{\bibfnamefont{M.}~\bibnamefont{Quandt}}, \bibnamefont{and}
  \bibinfo{author}{\bibfnamefont{H.}~\bibnamefont{Weigel}},
  \bibinfo{journal}{Lect.Notes Phys.} \textbf{\bibinfo{volume}{777}},
  \bibinfo{pages}{1} (\bibinfo{year}{2009}).

\bibitem[{\citenamefont{Graham et~al.}(2002)\citenamefont{Graham, Jaffe,
  Khemani, Quandt, Scandurra, and Weigel}}]{Graham:2002xq}
\bibinfo{author}{\bibfnamefont{N.}~\bibnamefont{Graham}},
  \bibinfo{author}{\bibfnamefont{R.~L.} \bibnamefont{Jaffe}},
  \bibinfo{author}{\bibfnamefont{V.}~\bibnamefont{Khemani}},
  \bibinfo{author}{\bibfnamefont{M.}~\bibnamefont{Quandt}},
  \bibinfo{author}{\bibfnamefont{M.}~\bibnamefont{Scandurra}},
  \bibnamefont{and} \bibinfo{author}{\bibfnamefont{H.}~\bibnamefont{Weigel}},
  \bibinfo{journal}{Nucl. Phys.} \textbf{\bibinfo{volume}{B645}},
  \bibinfo{pages}{49} (\bibinfo{year}{2002}).

\bibitem[{\citenamefont{Rajaraman}(1982)}]{Ra82}
\bibinfo{author}{\bibfnamefont{R.}~\bibnamefont{Rajaraman}},
  \emph{\bibinfo{title}{Solitons and Instantons}} (\bibinfo{publisher}{North
  Holland}, \bibinfo{year}{1982}).

\bibitem[{\citenamefont{Faulkner}(1977)}]{Faulkner:1977}
\bibinfo{author}{\bibfnamefont{J.~S.} \bibnamefont{Faulkner}},
  \bibinfo{journal}{J. Phys. C: Solid State Phys.}
  \textbf{\bibinfo{volume}{10}}, \bibinfo{pages}{4461} (\bibinfo{year}{1977}).

\bibitem[{\citenamefont{Graham et~al.}(2011)\citenamefont{Graham, Quandt, and
  Weigel}}]{Graham:2011fw}
\bibinfo{author}{\bibfnamefont{N.}~\bibnamefont{Graham}},
  \bibinfo{author}{\bibfnamefont{M.}~\bibnamefont{Quandt}}, \bibnamefont{and}
  \bibinfo{author}{\bibfnamefont{H.}~\bibnamefont{Weigel}},
  \bibinfo{journal}{Phys.Rev.} \textbf{\bibinfo{volume}{D84}},
  \bibinfo{pages}{025017} (\bibinfo{year}{2011}), 

\bibitem[{\citenamefont{Farhi et~al.}(2000)\citenamefont{Farhi, Graham, Jaffe,
  and Weigel}}]{Farhi:2000ws}
\bibinfo{author}{\bibfnamefont{E.}~\bibnamefont{Farhi}},
  \bibinfo{author}{\bibfnamefont{N.}~\bibnamefont{Graham}},
  \bibinfo{author}{\bibfnamefont{R.~L.} \bibnamefont{Jaffe}}, \bibnamefont{and}
  \bibinfo{author}{\bibfnamefont{H.}~\bibnamefont{Weigel}},
  \bibinfo{journal}{Nucl.Phys.} \textbf{\bibinfo{volume}{B585}},
  \bibinfo{pages}{443} (\bibinfo{year}{2000}), 

\bibitem[{\citenamefont{Graham et~al.}(2001)\citenamefont{Graham, Jaffe,
  Quandt, and Weigel}}]{Graham:2001dy}
\bibinfo{author}{\bibfnamefont{N.}~\bibnamefont{Graham}},
  \bibinfo{author}{\bibfnamefont{R.~L.} \bibnamefont{Jaffe}},
  \bibinfo{author}{\bibfnamefont{M.}~\bibnamefont{Quandt}}, \bibnamefont{and}
  \bibinfo{author}{\bibfnamefont{H.}~\bibnamefont{Weigel}},
  \bibinfo{journal}{Phys.Rev.Lett.} \textbf{\bibinfo{volume}{87}},
  \bibinfo{pages}{131601} (\bibinfo{year}{2001}), 

\bibitem[{\citenamefont{Dashen et~al.}(1974)\citenamefont{Dashen, Hasslacher,
  and Neveu}}]{Dashen:1974ci}
\bibinfo{author}{\bibfnamefont{R.~F.} \bibnamefont{Dashen}},
  \bibinfo{author}{\bibfnamefont{B.}~\bibnamefont{Hasslacher}},
  \bibnamefont{and} \bibinfo{author}{\bibfnamefont{A.}~\bibnamefont{Neveu}},
  \bibinfo{journal}{Phys.Rev.} \textbf{\bibinfo{volume}{D10}},
  \bibinfo{pages}{4114} (\bibinfo{year}{1974}).

\bibitem[{\citenamefont{Gervais et~al.}(1975)\citenamefont{Gervais, Jevicki,
  and Sakita}}]{Gervais:1975pa}
\bibinfo{author}{\bibfnamefont{J.-L.} \bibnamefont{Gervais}},
  \bibinfo{author}{\bibfnamefont{A.}~\bibnamefont{Jevicki}}, \bibnamefont{and}
  \bibinfo{author}{\bibfnamefont{B.}~\bibnamefont{Sakita}},
  \bibinfo{journal}{Phys.Rev.} \textbf{\bibinfo{volume}{D12}},
  \bibinfo{pages}{1038} (\bibinfo{year}{1975}).

\bibitem[{\citenamefont{Cahill et~al.}(1976)\citenamefont{Cahill, Comtet, and
  Glauber}}]{Cahill:1976im}
\bibinfo{author}{\bibfnamefont{K.~E.} \bibnamefont{Cahill}},
  \bibinfo{author}{\bibfnamefont{A.}~\bibnamefont{Comtet}}, \bibnamefont{and}
  \bibinfo{author}{\bibfnamefont{R.~J.} \bibnamefont{Glauber}},
  \bibinfo{journal}{Phys.Lett.} \textbf{\bibinfo{volume}{B64}},
  \bibinfo{pages}{283} (\bibinfo{year}{1976}).

\bibitem[{\citenamefont{Nakahara and Maki}(1982)}]{Nakahara:1982}
\bibinfo{author}{\bibfnamefont{M.}~\bibnamefont{Nakahara}} \bibnamefont{and}
  \bibinfo{author}{\bibfnamefont{K.}~\bibnamefont{Maki}},
  \bibinfo{journal}{Phys. Rev.} \textbf{\bibinfo{volume}{B25}},
  \bibinfo{pages}{7789} (\bibinfo{year}{1982}).

\bibitem[{\citenamefont{Graham and Jaffe}(1999)}]{Graham:1998qq}
\bibinfo{author}{\bibfnamefont{N.}~\bibnamefont{Graham}} \bibnamefont{and}
  \bibinfo{author}{\bibfnamefont{R.~L.} \bibnamefont{Jaffe}},
  \bibinfo{journal}{Nucl.Phys.} \textbf{\bibinfo{volume}{B544}},
  \bibinfo{pages}{432} (\bibinfo{year}{1999}), 

\bibitem[{\citenamefont{Flores-Hidalgo}(2002)}]{FloresHidalgo:2002at}
\bibinfo{author}{\bibfnamefont{G.}~\bibnamefont{Flores-Hidalgo}},
  \bibinfo{journal}{Phys.Lett.} \textbf{\bibinfo{volume}{B542}},
  \bibinfo{pages}{282} (\bibinfo{year}{2002}), 

\bibitem[{\citenamefont{Moussallam and Kalafatis}(1991)}]{Moussallam:1991rj}
\bibinfo{author}{\bibfnamefont{B.}~\bibnamefont{Moussallam}} \bibnamefont{and}
  \bibinfo{author}{\bibfnamefont{D.}~\bibnamefont{Kalafatis}},
  \bibinfo{journal}{Phys.Lett.} \textbf{\bibinfo{volume}{B272}},
  \bibinfo{pages}{196} (\bibinfo{year}{1991}).

\bibitem[{\citenamefont{Meier and Walliser}(1997)}]{Meier:1996ng}
\bibinfo{author}{\bibfnamefont{F.}~\bibnamefont{Meier}} \bibnamefont{and}
  \bibinfo{author}{\bibfnamefont{H.}~\bibnamefont{Walliser}},
  \bibinfo{journal}{Phys.Rept.} \textbf{\bibinfo{volume}{289}},
  \bibinfo{pages}{383} (\bibinfo{year}{1997}), 

\bibitem[{\citenamefont{Weigel}(2008)}]{Weigel:2008zz}
\bibinfo{author}{\bibfnamefont{H.}~\bibnamefont{Weigel}},
  \bibinfo{journal}{Lect.Notes Phys.} \textbf{\bibinfo{volume}{743}},
  \bibinfo{pages}{1} (\bibinfo{year}{2008}).

\bibitem[{\citenamefont{Rahi et~al.}(2009)\citenamefont{Rahi, Emig, Graham,
  Jaffe, and Kardar}}]{Rahi:2009hm}
\bibinfo{author}{\bibfnamefont{S.~J.} \bibnamefont{Rahi}},
  \bibinfo{author}{\bibfnamefont{T.}~\bibnamefont{Emig}},
  \bibinfo{author}{\bibfnamefont{N.}~\bibnamefont{Graham}},
  \bibinfo{author}{\bibfnamefont{R.~L.} \bibnamefont{Jaffe}}, \bibnamefont{and}
  \bibinfo{author}{\bibfnamefont{M.}~\bibnamefont{Kardar}},
  \bibinfo{journal}{Phys.Rev.} \textbf{\bibinfo{volume}{D80}},
  \bibinfo{pages}{085021} (\bibinfo{year}{2009}), 

\bibitem[{\citenamefont{Kenneth and Klich}(2006)}]{Kenneth06}
\bibinfo{author}{\bibfnamefont{O.}~\bibnamefont{Kenneth}} \bibnamefont{and}
  \bibinfo{author}{\bibfnamefont{I.}~\bibnamefont{Klich}},
  \bibinfo{journal}{Phys. Rev. Lett.} \textbf{\bibinfo{volume}{97}},
  \bibinfo{eid}{160401} (\bibinfo{year}{2006}).

\bibitem[{\citenamefont{Kenneth and Klich}(2008)}]{Kenneth:2007jk}
\bibinfo{author}{\bibfnamefont{O.}~\bibnamefont{Kenneth}} \bibnamefont{and}
  \bibinfo{author}{\bibfnamefont{I.}~\bibnamefont{Klich}},
  \bibinfo{journal}{Phys.Rev.} \textbf{\bibinfo{volume}{B78}},
  \bibinfo{pages}{014103} (\bibinfo{year}{2008}), 

\bibitem[{\citenamefont{Graham et~al.}(2013)\citenamefont{Graham, Quandt, and
  Weigel}}]{Graham:2013yza}
\bibinfo{author}{\bibfnamefont{N.}~\bibnamefont{Graham}},
  \bibinfo{author}{\bibfnamefont{M.}~\bibnamefont{Quandt}}, \bibnamefont{and}
  \bibinfo{author}{\bibfnamefont{H.}~\bibnamefont{Weigel}},
  \bibinfo{journal}{Phys.Lett.} \textbf{\bibinfo{volume}{B726}},
  \bibinfo{pages}{846} (\bibinfo{year}{2013}), 

\bibitem[{\citenamefont{Landau and Lifshitz}(1986)}]{Landau}
\bibinfo{author}{\bibfnamefont{L.~D.} \bibnamefont{Landau}} \bibnamefont{and}
  \bibinfo{author}{\bibfnamefont{E.~M.} \bibnamefont{Lifshitz}},
  \emph{\bibinfo{title}{Electrodynamics of Continuous Media}}
  (\bibinfo{publisher}{Butterworth--Heinemann}, \bibinfo{year}{1986}),
  chap.~\bibinfo{chapter}{80}.

\bibitem[{\citenamefont{Brevik and Milton}(2008)}]{Brevik:2008ry}
\bibinfo{author}{\bibfnamefont{I.~H.} \bibnamefont{Brevik}} \bibnamefont{and}
  \bibinfo{author}{\bibfnamefont{K.~A.} \bibnamefont{Milton}},
  \bibinfo{journal}{Phys.Rev.} \textbf{\bibinfo{volume}{E78}},
  \bibinfo{pages}{011124} (\bibinfo{year}{2008}), 

\bibitem[{\citenamefont{Milton et~al.}(2010)\citenamefont{Milton, Wagner,
  Parashar, and Brevik}}]{Milton:2010yw}
\bibinfo{author}{\bibfnamefont{K.~A.} \bibnamefont{Milton}},
  \bibinfo{author}{\bibfnamefont{J.}~\bibnamefont{Wagner}},
  \bibinfo{author}{\bibfnamefont{P.}~\bibnamefont{Parashar}}, \bibnamefont{and}
  \bibinfo{author}{\bibfnamefont{I.~H.} \bibnamefont{Brevik}},
  \bibinfo{journal}{Phys.Rev.} \textbf{\bibinfo{volume}{D81}},
  \bibinfo{pages}{065007} (\bibinfo{year}{2010}), 

\bibitem[{\citenamefont{Rosa et~al.}(2010)\citenamefont{Rosa, Dalvit, and
  Milonni}}]{PhysRevA.81.033812}
\bibinfo{author}{\bibfnamefont{F.~S.~S.} \bibnamefont{Rosa}},
  \bibinfo{author}{\bibfnamefont{D.~A.~R.} \bibnamefont{Dalvit}},
  \bibnamefont{and} \bibinfo{author}{\bibfnamefont{P.~W.}
  \bibnamefont{Milonni}}, \bibinfo{journal}{Phys. Rev. A}
  \textbf{\bibinfo{volume}{81}}, \bibinfo{pages}{033812}
  (\bibinfo{year}{2010}).

\bibitem[{\citenamefont{Rosa et~al.}(2011)\citenamefont{Rosa, Dalvit, and
  Milonni}}]{PhysRevA.84.053813}
\bibinfo{author}{\bibfnamefont{F.~S.~S.} \bibnamefont{Rosa}},
  \bibinfo{author}{\bibfnamefont{D.~A.~R.} \bibnamefont{Dalvit}},
  \bibnamefont{and} \bibinfo{author}{\bibfnamefont{P.~W.}
  \bibnamefont{Milonni}}, \bibinfo{journal}{Phys. Rev. A}
  \textbf{\bibinfo{volume}{84}}, \bibinfo{pages}{053813}
  (\bibinfo{year}{2011}).

\bibitem[{\citenamefont{Newton}(1982)}]{Newton:1982qc}
\bibinfo{author}{\bibfnamefont{R.~G.} \bibnamefont{Newton}},
  \emph{\bibinfo{title}{Scattering Theory of Waves and Particles}}
  (\bibinfo{publisher}{Springer, New York}, \bibinfo{year}{1982}).

\bibitem[{\citenamefont{Chadan and Sabatier}(1977)}]{Chadan:1977pq}
\bibinfo{author}{\bibfnamefont{K.}~\bibnamefont{Chadan}} \bibnamefont{and}
  \bibinfo{author}{\bibfnamefont{P.}~\bibnamefont{Sabatier}},
  \emph{\bibinfo{title}{{Inverse Problems in Quantum Scattering Theory}}}
  (\bibinfo{publisher}{Springer New York}, \bibinfo{year}{1977}).

\bibitem[{\citenamefont{Weinberg}(1995)}]{Weinberg:1995mt}
\bibinfo{author}{\bibfnamefont{S.}~\bibnamefont{Weinberg}},
  \emph{\bibinfo{title}{{The Quantum Theory of Fields. Vol. 1: Foundations}}}
  (\bibinfo{publisher}{Cambridge University Press}, \bibinfo{year}{1995}).

\bibitem[{\citenamefont{Ring and Schuck}(1980)}]{Ri80}
\bibinfo{author}{\bibfnamefont{P.}~\bibnamefont{Ring}} \bibnamefont{and}
  \bibinfo{author}{\bibfnamefont{P.}~\bibnamefont{Schuck}},
  \emph{\bibinfo{title}{The Nuclear Many Body Problem}}
  (\bibinfo{publisher}{Springer}, \bibinfo{year}{1980}).

\bibitem[{\citenamefont{Callan and Klebanov}(1985)}]{Callan:1985hy}
\bibinfo{author}{\bibfnamefont{C.~G.} \bibnamefont{Callan}} \bibnamefont{and}
  \bibinfo{author}{\bibfnamefont{I.~R.} \bibnamefont{Klebanov}},
  \bibinfo{journal}{Nucl.Phys.} \textbf{\bibinfo{volume}{B262}},
  \bibinfo{pages}{365} (\bibinfo{year}{1985}).

\bibitem[{\citenamefont{Callan et~al.}(1988)\citenamefont{Callan, Hornbostel,
  and Klebanov}}]{Callan:1987xt}
\bibinfo{author}{\bibfnamefont{C.~G.} \bibnamefont{Callan}},
  \bibinfo{author}{\bibfnamefont{K.}~\bibnamefont{Hornbostel}},
  \bibnamefont{and} \bibinfo{author}{\bibfnamefont{I.~R.}
  \bibnamefont{Klebanov}}, \bibinfo{journal}{Phys.Lett.}
  \textbf{\bibinfo{volume}{B202}}, \bibinfo{pages}{269} (\bibinfo{year}{1988}).

\bibitem[{\citenamefont{Blaizot et~al.}(1988)\citenamefont{Blaizot, Rho, and
  Scoccola}}]{Blaizot:1988ge}
\bibinfo{author}{\bibfnamefont{J.~P.} \bibnamefont{Blaizot}},
  \bibinfo{author}{\bibfnamefont{M.}~\bibnamefont{Rho}}, \bibnamefont{and}
  \bibinfo{author}{\bibfnamefont{N.~N.} \bibnamefont{Scoccola}},
  \bibinfo{journal}{Phys.Lett.} \textbf{\bibinfo{volume}{B209}},
  \bibinfo{pages}{27} (\bibinfo{year}{1988}).

\bibitem[{\citenamefont{Scoccola and Walliser}(1998)}]{Scoccola:1998eq}
\bibinfo{author}{\bibfnamefont{N.~N.} \bibnamefont{Scoccola}} \bibnamefont{and}
  \bibinfo{author}{\bibfnamefont{H.}~\bibnamefont{Walliser}},
  \bibinfo{journal}{Phys.Rev.} \textbf{\bibinfo{volume}{D58}},
  \bibinfo{pages}{094037} (\bibinfo{year}{1998}).

\bibitem[{\citenamefont{Panofsky and Phillips}(1962)}]{Panofsky}
\bibinfo{author}{\bibfnamefont{W.~K.~H.} \bibnamefont{Panofsky}}
  \bibnamefont{and} \bibinfo{author}{\bibfnamefont{M.}~\bibnamefont{Phillips}},
  \emph{\bibinfo{title}{Classical Electricity and Magnetism}}
  (\bibinfo{publisher}{Addison--Wesley, Reading, MA.}, \bibinfo{year}{1962}),
  chap.~\bibinfo{chapter}{6}.

\bibitem[{\citenamefont{Jackson}(1975)}]{Jackson}
\bibinfo{author}{\bibfnamefont{J.~D.} \bibnamefont{Jackson}},
  \emph{\bibinfo{title}{Classical Electrodynamics}}
  (\bibinfo{publisher}{Wilely, New York}, \bibinfo{year}{1975}),
  chap.~\bibinfo{chapter}{4}.

\bibitem[{\citenamefont{Mie}(1908)}]{Mie:1908}
\bibinfo{author}{\bibfnamefont{G.}~\bibnamefont{Mie}}, \bibinfo{journal}{Ann.
  Phys.} \textbf{\bibinfo{volume}{25}}, \bibinfo{pages}{377}
  (\bibinfo{year}{1908}).

\bibitem[{\citenamefont{Tsao}(1993)}]{Tsao:1993}
\bibinfo{author}{\bibfnamefont{P.-H.} \bibnamefont{Tsao}},
  \bibinfo{journal}{Am. J. Phys.} \textbf{\bibinfo{volume}{61}},
  \bibinfo{pages}{823} (\bibinfo{year}{1993}).

\bibitem[{\citenamefont{Schwinger et~al.}(1998)\citenamefont{Schwinger, Milton,
  DeRaad~Jr., and Tsai}}]{Schwinger}
\bibinfo{author}{\bibfnamefont{J.~S.} \bibnamefont{Schwinger}},
  \bibinfo{author}{\bibfnamefont{K.~A.} \bibnamefont{Milton}},
  \bibinfo{author}{\bibfnamefont{L.~L.} \bibnamefont{DeRaad~Jr.}},
  \bibnamefont{and} \bibinfo{author}{\bibfnamefont{W.}~\bibnamefont{Tsai}},
  \emph{\bibinfo{title}{Classical Electrodynamics}}
  (\bibinfo{publisher}{Perseus/Westview}, \bibinfo{year}{1998}),
  chap.~\bibinfo{chapter}{7}.

\end{thebibliography}

\end{document}